\documentclass[]{jfm}

\usepackage{graphicx}
\graphicspath{{Figs/}}
\usepackage{newtxtext}
\usepackage{newtxmath}
\usepackage{natbib}
\usepackage{hyperref}
\usepackage{array}					
\usepackage[graphicx]{realboxes}	
\usepackage{comment}

\newcommand{\vect}[1]{\boldsymbol{#1}}
\newcommand{\DCf}{\Delta C_f}	
\newcommand{\DCd}{\Delta C_D}	
\newcommand{\DE}{\Delta E}      
\newcommand{\DP}{\Delta P_n}	

\title[Aerodynamics of a transonic profile with spanwise forcing]{The aerodynamic performance of a transonic airfoil with spanwise forcing}

\author{Niccolò Berizzi
\aff{1},
Davide Gatti
\aff{2},
Giulio Soldati
\aff{3},
Sergio Pirozzoli
\aff{3}
\and 
Maurizio Quadrio
\aff{1}
\corresp{\email{maurizio.quadrio@polimi.it}}
}
\affiliation{
\aff{1} Dipartimento di Scienze e Tecnologie Aerospaziali, Politecnico di Milano,
Via La Masa 34, 20156 Milano, Italy \\
\aff{2} Institute of Fluid Mechanics, Karlsruhe Institute of Technology,
    Kaiserstra\ss e 10, 76131 Karlsruhe, Germany \\
\aff{3} Dipartimento di Meccanica e Aeronautica, Università di Roma "La Sapienza", Via Eudossiana 18, 00184 Roma, Italy
}

\begin{document}
\maketitle

\begin{abstract}
Spanwise wall forcing in the form of streamwise-travelling waves is applied 
to the suction side of a transonic airfoil with a shock wave to reduce aerodynamic drag. 
The study, conducted using direct numerical simulations, extends earlier findings by Quadrio {\em et al}. ({\em J. Fluid Mech.} vol. 942, 2022, R2) 
and confirms that the wall manipulation shifts the shock wave on the suction side 
towards the trailing edge of the profile, thereby enhancing its aerodynamic efficiency.
A parametric study over the parameters of wall forcing is carried out for the Mach number set at 0.7 and the Reynolds number at 300,000. 
Similarities and differences with the incompressible plane case are discussed;  
for the first time, we describe how the interaction between the shock wave 
and the boundary layer is influenced by flow control via spanwise forcing.
With suitable combinations of control parameters, the shock is delayed, 
and results in a separated region whose length correlates well with friction reduction. 
The analysis of the transient process following the sudden application of control 
is used to link flow separation with the intensification of the shock wave.
\end{abstract}


\section{Introduction}
\label{sec:introduction}

Environmental pollution and global warming driven by $\mathrm{CO_2}$ emissions are severe 
global concerns; the civil aeronautical sector is a significant contributor.
The United Nations' International Civil Aviation Organization (ICAO) forecasts a triplication of the aviation emissions by 2050, accounting for 25\% of the global carbon budget \citep{graver-zhang-rutherford-2019}.
Hence, the quest to reduce atmospheric pollutants, alongside pressing economic reasons, motivates the industry efforts towards more efficient vehicles.
In air transportation, the roughly linear dependency of fuel consumption on aerodynamic resistance drives research into flow control methods for 
drag reduction.
However, controlling turbulence in flow conditions typical of transport applications presents many scientific and engineering challenges. 
In mature applications involving fluid flows, drag --- particularly skin-friction drag, which is intimately related to the turbulent nature of high-Reynolds-number flows --- is difficult to abate.

\begin{figure}
\centering
\includegraphics{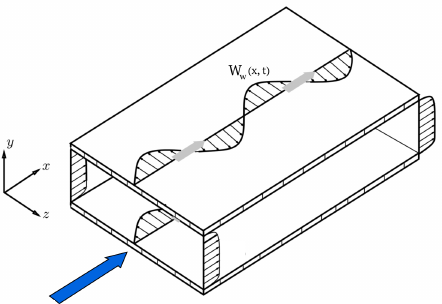}
\caption{Sketch of wall manipulation through streamwise-travelling waves of 
spanwise velocity in a channel flow setup. Figure adapted from \cite{gatti-quadrio-2016}.}
\label{fig:StTW}
\end{figure}

Flow control techniques aimed at friction reduction are usually 
classified into passive and active. The latter require extra energy 
but can yield higher net energetic benefits. Among them, spanwise 
forcing based on a space- and/or time-dependent distribution of 
spanwise velocity at the wall stands out for its energy-saving 
potential. Several implementations of spanwise forcing were
considered over the years, starting from the simplest spanwise-oscillating 
wall, introduced by \cite{jung-mangiavacchi-akhavan-1992} and thoroughly 
studied by \cite{quadrio-ricco-2004}, in which the spanwise motion of the wall 
is uniform in space and varies harmonically in time.
The most promising technique of this class is the combined space-time 
version, i.e., the streamwise-travelling waves of spanwise velocity 
(hereinafter, StTW), introduced years later by 
\cite{quadrio-ricco-viotti-2009}. This technique involves manipulating 
the wall velocity $W_w$ according to 
\begin{equation}
W_w(x, t) = A \sin \left( \kappa_x x - \omega t \right) ,
\label{eq:bc4sttw}
\end{equation}
where $x$ and $t$ are the streamwise coordinate and time, respectively,
while $A$, $\kappa_x$, and $\omega$ are the control amplitude, 
and the spatial and temporal frequencies of the forcing, 
respectively (see figure \ref{fig:StTW} for a sketch).
\cite{quadrio-ricco-2011} qualitatively explained the working
mechanism of the StTW: the waves generate a transversal Stokes layer (whose thickness is determined by the control parameters), which
interacts favorably with the near-wall turbulence. The effectiveness of StTW has received
experimental verification: \cite{auteri-etal-2010} measured up to $33\%$
drag reduction in a pipe flow by spatially discretizing the sinusoidal
waveform with independently rotating pipe slices; \cite{bird-santer-morrison-2018}
developed a planar actuator for StTW, consisting of a tensioned
membrane skin mounted on a kagome lattice; \cite{marusic-etal-2021}
and \cite{knoop-etal-2024} extended the spatially discrete actuator of \cite{auteri-etal-2010} to the planar geometry. Finally, the recent contribution by \cite{gallorini-quadrio-2024} has led to a satisfactory understanding of the discrete form of the forcing.

These promising results mostly derive from studies of incompressible
flows in simple planar geometries, and at relatively low Reynolds numbers
($\Rey$). While the $\Rey$-dependence has been thoroughly discussed
by \cite{gatti-quadrio-2016}, flow compressibility has been put into focus more recently. \cite{yao-hussain-2019} found that
compressibility lead to additional drag reduction in a
supersonic channel with oscillating walls. However, \cite{gattere-etal-2024}, besides considering the general StTW case, reviewed the procedure to compare controlled and uncontrolled cases, and concluded that the
power budget of StTW in planar geometry is essentially unaffected
by flow compressibility, once the comparison is properly conducted.
Nevertheless, StTW have rarely been analyzed in the compressible regime
in combination with non-planar walls, despite this is the typical
application scenario in aeronautics, where shock waves can develop, and
pressure and friction contributions are comparable in the global
drag budget \citep{abbas-devicente-valero-2013}. 

The idea that control
techniques for friction drag reduction might affect pressure drag was
put forward by \cite{mele-tognaccini-catalano-2016}, who performed
Reynolds-averaged Navier--Stokes (RANS) simulations around a transonic
wing-body configuration with riblets installed. \cite{banchetti-luchini-quadrio-2020}
verified via Direct Numerical Simulations (DNS) that this is indeed
the case for StTW in an incompressible flow over a non-planar wall.
Finally, \cite{quadrio-etal-2022} applied StTW in a physically richer scenario,
relaxing both the planarity and incompressibility assumptions. They employed
DNS to describe the turbulent flow over a modern transonic airfoil
where a shock wave develops. StTW were applied to a portion of the
suction side of the wing, and their effects were measured in terms of
global lift and drag forces, together with the evolution of friction and
pressure along the chord.
They observed that StTW affect both friction and pressure distributions, with the most
significant flow modifications ensuing from changes in the position
and intensity of the shock wave. Specifically, a more intense and
delayed shock was observed, leading to a significant improvement in the
aerodynamic efficiency of the wing. At the lower incidence that
re-establishes the required lift with the control enabled, a net reduction of $9\%$ was estimated for the drag of the entire aircraft.

In this work we extend these results, and seek a deeper physical understanding.
We keep using StTW forcing, that is interesting because of its large local drag reduction and net savings, which make the global effects more self-evident; its hassle-free implementation in a simulation via boundary conditions is attractive.
The original study of \cite{quadrio-etal-2022} only tested two StTW configurations, 
where the control parameters were selected on the basis of incompressible information; 
the extent and position of the active surface area were deduced empirically.
Moreover, a characterization of the physics behind the control-induced flow changes was not attempted. 
In this study, we intend neither to carry out a full optimization of StTW for this particular airfoil, nor to assess the effects of compressibility alone, for which the picture described by \cite{gattere-etal-2024} is exhaustive enough for the planar geometry. Instead, we set two goals. The first is to explore the parameter space defined by the control variables $(A, \kappa_x, \omega)$ and augmented by the position and extent of the actuated region, to identify analogies and differences with the plane channel flow. The second is to describe how the control alters the complex physics of the flow, characterized by the mutual interaction of the shock wave and the turbulent boundary layer.

The paper is structured as follows: in \S\ref{sec:numerical}, we describe
the flow configuration and the numerical tools adopted in the analysis;
in \S\ref{sec:aerodynamics}, the aerodynamic performance of
the system subject to spanwise forcing is studied. In \S\ref{sec:flowphys}, we
address the flow physics, focusing first on the evolution of the boundary
layer along the chord in \S\ref{sec:evolution}, and then on the response
of the flow after the sudden imposition of the forcing in \S\ref{sec:transient}.
Finally, a concluding discussion is presented in \S\ref{sec:conclusions}.

\section{Numerical methods and procedures}
\label{sec:numerical}

\subsection{DNS solver and computational setup} 

The DNS solver used in this study is the same employed by
\cite{quadrio-etal-2022}, and was extensively described
and validated earlier by \cite{memmolo-bernardini-pirozzoli-2018}.
The compressible Navier--Stokes equations are solved for an ideal gas,
with the heat flux vector and the viscous stress tensor modeled by
Fourier's law and the Newtonian hypothesis, respectively. The dependence
of dynamic viscosity $\mu$ on temperature $T$ is described by the power law
$\mu / \mu_\infty = \left( T / T_{\infty} \right)^{0.76}$
\citep{smits-dussauge-2006}. Here and throughout the paper,
the subscript ‘$\infty$’ refers to the freestream undisturbed flow.
The equations, cast in integral form,
are discretized using a second-order, energy-consistent finite-volumes
method \citep{pirozzoli-2011}. In the presence of shock waves, detected
by a modified Ducros sensor \citep{ducros-etal-1999}, the code switches
locally to a third-order weighted essentially non-oscillatory scheme
\citep{liu-osher-chan-1994}.

The case of interest is the same considered in \cite{quadrio-etal-2022} \emph{i.e.}, an airfoil immersed in a uniform flow. The sectional profile
is the supercritical V2C airfoil, designed within the European project
TFAST \citep{doerffer-etal-2021}, and already studied experimentally \citep{placek-ruchala-2018}
and numerically \citep{szubert-etal-2016, zauner-detullio-sandham-2019}.
The angle of attack is $\alpha = 4^\circ$, which provides the maximum
lift-to-drag ratio. The airflow has inflow
conditions of Mach number $M_\infty \equiv U_\infty/a_\infty = 0.7$
(where $U_\infty$ is the freestream velocity, and $a_\infty$ is the speed
of sound in the free stream) and Reynolds number $\Rey_\infty \equiv
U_\infty c / \nu_\infty = 3 \times 10^5$ (where $c$ is the profile chord,
and $\nu_\infty$ is the kinematic viscosity in the free stream). Periodicity is enforced
in the spanwise direction, and the inflow is laminar: to avoid the large excursions of the transition location \citep{zauner-detullio-sandham-2019}, the boundary layer is tripped, as in experiments. A small Gaussian blob of wall-normal random volume force \citep{schlatter-orlu-2012} acting on both sides of the airfoil $0.1 c$ downstream of the leading edge.
The intensity of the tripping force ($100 U_\infty^2/c$) was selected as the minimum value granting healthy development of the turbulent boundary layer.

Unless otherwise indicated, the length scale $c$ and velocity scale 
$U_\infty$ are used to form dimensionless quantities, and are omitted hereinafter.
Alternatively, viscous units will also be adopted, where the 
reference velocity is the friction velocity ($u_\tau=\sqrt{\tau_w/\rho}$), 
and the reference length is the viscous length ($\nu/u_\tau$). 
In these definitions, $\tau_w, \rho, \nu$ are the wall-shear stress, 
density, and kinematic viscosity, respectively. 
Quantities made non-dimensional 
with the viscous units extracted from the reference uncontrolled simulation are 
denoted with a `+' superscript, while the `$\ast$' scaling denotes quantites made non-dimensional with the friction velocity of the current simulation.
Note, however, 
that the wall-shear stress varies significantly along the profile. 
In this paper, friction is evaluated on the suction side at $x=0.4$, 
a position that is downstream enough for the actuation to actually 
develop its effects, but also sufficiently upstream of the shock 
such that the wall pressure gradient is almost constant across the 
various simulations. 
This aspect is further discussed in \S\ref{sec:parameters} and 
in Appendix \ref{app:friction}. 

The computational setup replicates the baseline case described by
\cite{quadrio-etal-2022}, who validated it in terms of resolution
requirements. The mesh features a C-type topology with $4096 \times 512 \times 256$ cells. 
It has a radius $25$ chords and a spanwise extension of $L_z = 0.1$, which is sufficiently large to grant that the flow statistics are independent from this length.
For verification, an additional simulation was carried out for the reference flow with a spanwise extension of $L_z = 0.4$, only to find overlapping results.
The cells are uniformly spaced in the spanwise direction, while a hyperbolic-tangent
clustering is adopted in the wall-normal direction. The mesh meets the
requirements for a fully-resolved DNS, with $\Delta x^+ < 10$,
$\Delta y^+ < 0.5$, and $\Delta z^+ < 5$ at the wall, where $\Delta x$,
$\Delta y$, and $\Delta z$ are the grid spacings in the $x$, $y$, and
$z$ directions, respectively. Curvilinear coordinates $\xi$ and $\eta$
will be used in the following to denote directions tangent and normal to the wall,
respectively. Time integration is performed using a low-storage,
third-order Runge--Kutta scheme, with a fixed time step of $\Delta t =
1.2 \times 10^{-4}$, chosen to keep the maximum Courant–-Friedrichs-–Lewy
number below unity. 
The quantities of interest are averaged in the
spanwise direction and over time for a minimum duration of $\Delta T = 40$;
the most significant cases, discussed in the next section, are run for
a maximum $\Delta T = 130$. 
The initial transient, determined on a case-by-case basis, is excluded from the time-averaging process.
An {\em a posteriori} analysis of the lift and drag time histories using the algorithm developed by \cite{russo-luchini-2017} confirms that the ratio between the standard deviation $\sigma$
of the estimated mean value of the lift coefficient and the coefficient's value remains below $0.4\%$ across all cases; the same quantity for the drag coefficient is below $0.7\%$. These metrics are presented as $\sigma_L$ and $\sigma_D$ in table \ref{tab:parameters}.

\begin{figure}
\centering
\includegraphics[width=\textwidth]{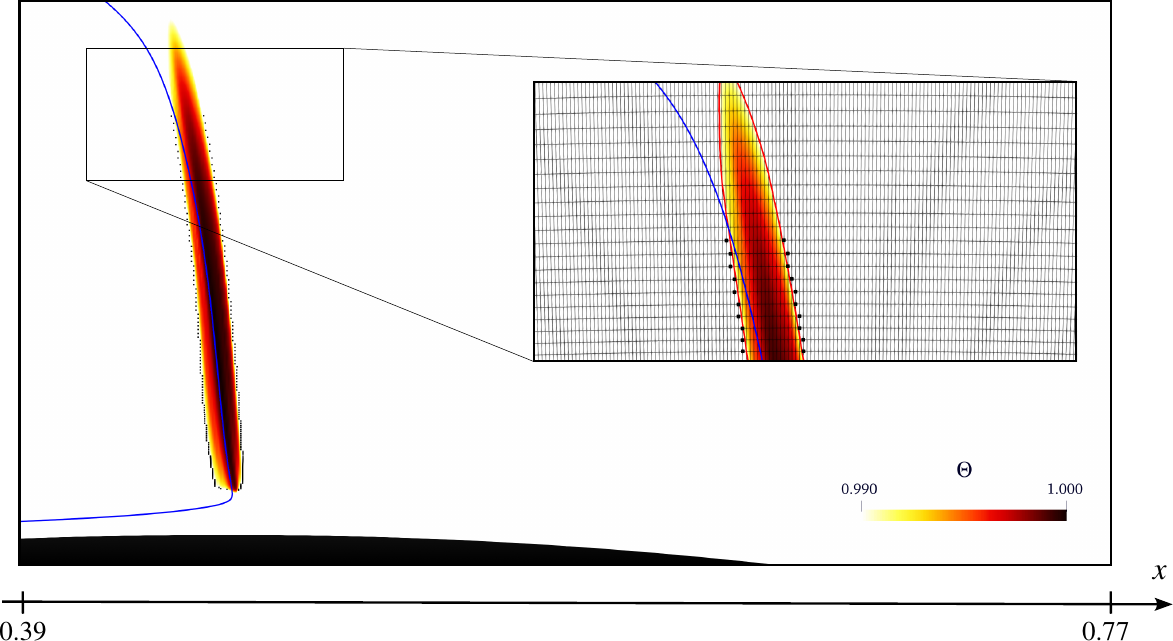}
\caption{Visualization of the mean modified Ducros shock sensor over the suction side of the airfoil (reference simulation), highlighting regions where $\Theta>0.99$. The blue line is the isoline at $M=1$, and the black dots denote the cells used to compute the shock intensity. The inset shows the upper edge of the shock superimposed to the grid, with the red isoline corresponding to $\Theta=0.99$.}
\label{fig:ducros}
\end{figure}

The discussion that follows will consider the position and strength of the shock wave. 
To identify the shock, we use the modified Ducros sensor \citep{pirozzoli-2011}
\begin{equation}
\Theta = \max \left( - \frac{\vect{\nabla} \vect{\cdot} \vect{u}}
{\sqrt{\left(\vect{\nabla} \vect{\cdot} \vect{u} \right)^2 + \left(\vect{\nabla} \times \vect{u}\right)^2 + \left( U_\infty / c \right)^2}}; 0 \right)
\end{equation}
where $\vect{u}$ is the local velocity vector. Figure~\ref{fig:ducros} displays the mean value of
$\Theta$ near the shock wave for the reference case.
We define a point to be within the shock if $\Theta > 0.99$. The shock position, denoted with $x_s$ in table \ref{tab:parameters}, is computed as the $x$ coordinate of the point closest to the wall where $\Theta > 0.99$.
The shock intensity, denoted as $I$ in table \ref{tab:parameters}, is determined
as the wall-normal average of the ratio between the pre- and post-shock Mach numbers. At a given wall distance, the pair of Mach number values is extracted at the two sides of the $\Theta = 0.99$ iso-line, i.e. at the black dots of figure~\ref{fig:ducros}.

\subsection{Description of the numerical experiments}

The streamwise-travelling waves \eqref{eq:bc4sttw} are characterized by five parameters. 
Three of them also apply to the canonical case of plane channel, namely
the spatial wavenumber $\kappa_x$, the pulsation $\omega$, 
and the amplitude $A$. Additionally, here the finite 
controlled region on the suction side of the airfoil begins at the
position $x_b$ and ends at the position $x_e$. Following \cite{yudhistira-skote-2011}, 
an exponential smoothing function acting on a length scale of $0.05$ chord units
is used at the boundary between active and inactive regions.
%
%
The space of these five parameters is explored with twenty-nine DNS, listed in table 
\ref{tab:parameters}, where the control parameters are reported along
with some global results to be discussed later. 

\begin{figure}
\centering
\includegraphics{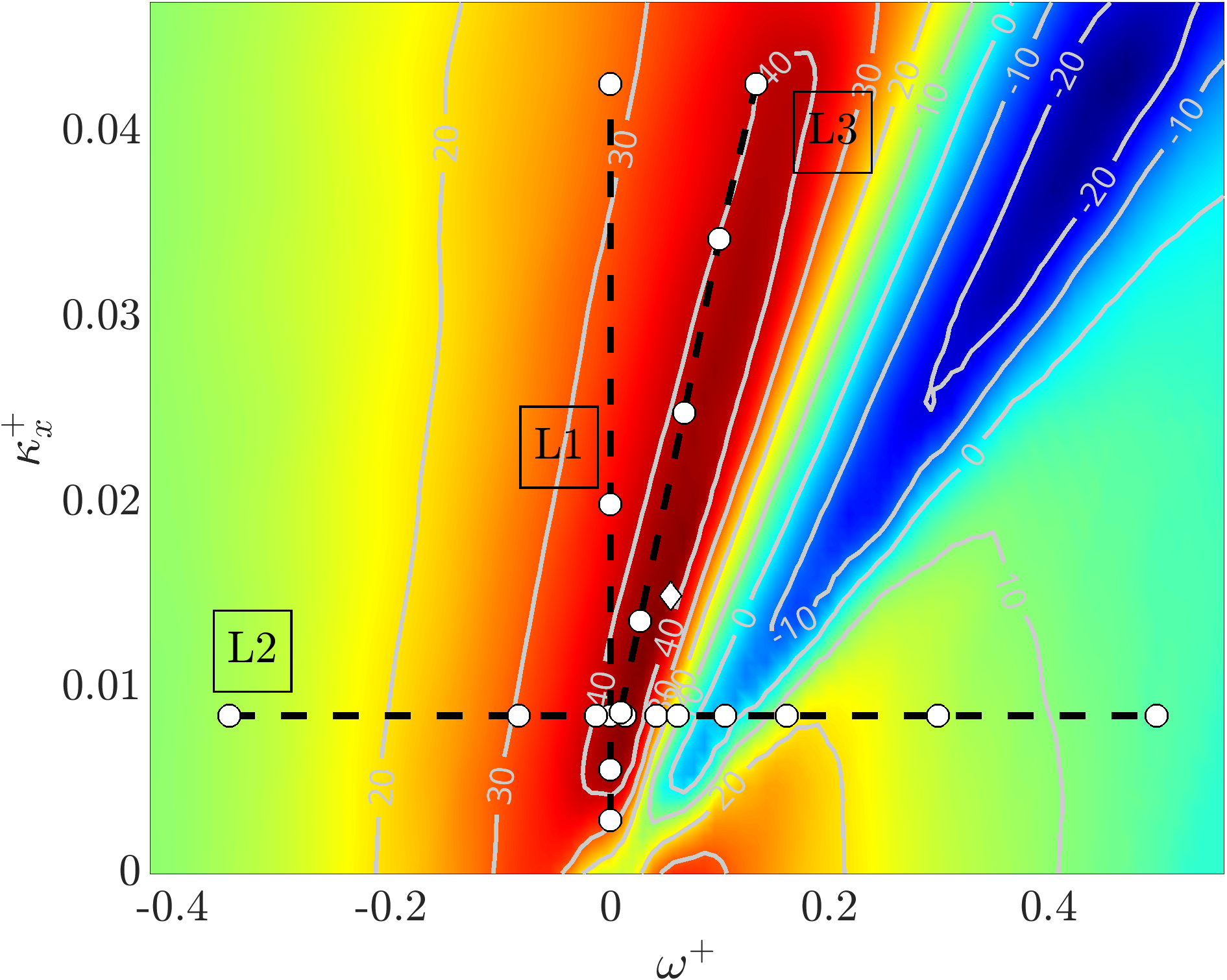}
\caption{Control parameters for flow cases cases C1--C20, superposed to the drag reduction map for incompressible plane channel flow \citep{gatti-quadrio-2016}. 
The diamond corresponds to case C2 of \cite{quadrio-etal-2022}. The isolines indicate percentage reduction of friction drag.}
\label{fig:paramspace}
\end{figure}

Case REF corresponds to the baseline unmanipulated flow at incidence angle of $4^\circ$.
Cases from C1 to C20 have a fixed amplitude ($A=0.684$, equivalent to $A^+=11$), 
and fixed starting point ($x_b=0.2$) and end point ($x_e=0.78$) of the forcing, 
as in \cite{quadrio-etal-2022}. 
These cases are depicted with dots in figure 
\ref{fig:paramspace}, where the drag reduction map for the 
incompressible channel flow at $\Rey_\tau=200$ 
\citep[][]{gatti-quadrio-2016} is reported. 
The first subset of cases lies on the vertical line L1 at 
$\omega=0$, corresponding to steady forcing; the second subset lies on the 
horizontal line L2 at constant wavelength, crossing both 
drag-reducing and drag-increasing regions of the parameter space; 
the third subset lies 
along the oblique line L3, corresponding to the ridge of maximum drag 
reduction in incompressible channel flow.
Additional cases from C21 to C26 are designed to understand 
the effects of the remaining control parameters ($x_b$, $x_e$ and $A$).
One simulation (labelled as R10 in table \ref{tab:parameters}) modifies one of 
the best-performing cases (namely C10) by reducing the incidence angle 
from $4^\circ$ to $3.45^\circ$, in such a way that the controlled airfoil yields
the same lift of the original uncontrolled case. Case RREF is the 
corresponding unmanipulated simulation at reduced incidence.

In the remainder of the paper, special attention will be paid to
cases C10 (one of the best performers), C12 (the only case yielding drag increase), 
and R10 (same as C10 but with reduced incidence).
For readability, they will be renamed DR (drag reduction), DI (drag-increase) and DRCL (drag reduction at constant lift).

\begin{table}
\Rotatebox{90}{
\centering
\setlength{\extrarowheight}{0.08cm}
\begin{tabular}{l|rrrrr|crrrrrrcrc|rrrr}
Case ID & $\kappa_x$ & $\omega$ & $A$ & $x_b$ & $x_e$ & $C_L$ & $\sigma_{L}$ & $C_D$ & $\sigma_{D}$ & $C_{D,f}$ & $C_{D,p}$ & $C_M$ & $E$ & $x_s$ & $I$ & $\DCf$ & $\DCd$ & $\DE$ & $\DP$\\
\hline
REF & $-$ & $-$ & $-$ & $-$ & $-$ & 0.747 & 0.12 & 0.0241 & 0.21 & 0.008 & 0.016 & -0.512 & 31.0 & 0.47 & 1.12 & $-$ & $-$ & $-$ & $-$ \\
RREF & $-$ & $-$ & $-$ & $-$ & $-$ & 0.664 & 0.20 & 0.0221 & 0.20 & 0.009 & 0.013 & -0.446 & 30.0 & 0.47 & 1.03 & $-$ & $-$ & $-$ & $-$ \\
C1 & 36.9 & 0.0 & 0.684 & 0.2 & 0.78 & 0.846 & 0.33 & 0.0250 & 0.39 & 0.007 & 0.018 & -0.570 & 33.9 & 0.52 & 1.11 & -37.3 & 3.5 & 9.5 & -2.2 \\
C2 & 71.7 & 0.0 & 0.684 & 0.2 & 0.78 & 0.826 & 0.12 & 0.0243 & 0.53 & 0.007 & 0.017 & -0.558 & 34.0 & 0.51 & 1.11 & -46.1 & 0.6 & 9.8 & -2.8 \\
C3 & 108.7 & 0.0 & 0.684 & 0.2 & 0.78 & 0.826 & 0.31 & 0.0242 & 0.47 & 0.007 & 0.017 & -0.558 & 34.1 & 0.51 & 1.09 & -40.6 & 0.5 & 10.0 & -2.5 \\
C4 & 254.2 & 0.0 & 0.684 & 0.2 & 0.78 & 0.827 & 0.37 & 0.0243 & 0.55 & 0.007 & 0.017 & -0.559 & 34.1 & 0.51 & 1.11 & -30.5 & 0.7 & 9.9 & -1.1 \\
C5 & 543.1 & 0.0 & 0.684 & 0.2 & 0.78 & 0.826 & 0.19 & 0.0242 & 0.44 & 0.007 & 0.017 & -0.559 & 34.1 & 0.51 & 1.15 & -20.3 & 0.3 & 10.2 & 0.4 \\
C6 & 108.6 & -332.2 & 0.684 & 0.2 & 0.78 & 0.788 & 0.29 & 0.0239 & 0.34 & 0.008 & 0.016 & -0.537 & 32.9 & 0.49 & 1.13 & -9.2 & -0.9 & 6.3 & 12.8 \\
C7 & 108.6 & -79.8 & 0.684 & 0.2 & 0.78 & 0.854 & 0.25 & 0.0249 & 0.35 & 0.007 & 0.018 & -0.575 & 34.4 & 0.53 & 1.16 & -23.7 & 3.1 & 10.9 & 2.7 \\
C8 & 108.6 & -12.7 & 0.684 & 0.2 & 0.78 & 0.834 & 0.21 & 0.0244 & 0.74 & 0.007 & 0.017 & -0.562 & 34.2 & 0.52 & 1.12 & -37.0 & 1.2 & 10.3 & -1.6 \\
C9 & 108.6 & 12.7 & 0.684 & 0.2 & 0.78 & 0.827 & 0.26 & 0.0243 & 0.33 & 0.007 & 0.017 & -0.559 & 34.0 & 0.51 & 1.11 & -44.4 & 0.9 & 9.6 & -2.5 \\
C10/DR & 108.6 & 39.8 & 0.684 & 0.2 & 0.78 & 0.838 & 0.19 & 0.0246 & 0.23 & 0.007 & 0.017 & -0.565 & 34.1 & 0.52 & 1.13 & -40.8 & 1.8 & 10.1 & -1.7 \\
R10/DRCL & 108.6 & 39.8 & 0.684 & 0.2 & 0.78 & 0.750 & 0.24 & 0.0211 & 0.17 & 0.008 & 0.013 & -0.498 & 35.5 & 0.52 & 1.08 & -36.5 & -4.4 & 18.3 & $-$ \\
C11 & 108.6 & 58.7 & 0.684 & 0.2 & 0.78 & 0.797 & 0.18 & 0.0241 & 0.40 & 0.008 & 0.016 & -0.541 & 33.0 & 0.50 & 1.14 & -19.8 & 0.1 & 6.5 & 2.2 \\
C12/DI & 108.6 & 99.6 & 0.684 & 0.2 & 0.78 & 0.738 & 0.13 & 0.0244 & 0.45 & 0.008 & 0.016 & -0.506 & 30.3 & 0.46 & 1.12 & 26.1 & 1.1 & -2.3 & 14.0 \\
C13 & 108.6 & 153.4 & 0.684 & 0.2 & 0.78 & 0.787 & 0.16 & 0.0238 & 0.36 & 0.008 & 0.016 & -0.536 & 33.1 & 0.49 & 1.16 & -11.1 & -1.3 & 6.8 & 7.5 \\
C14 & 108.6 & 285.2 & 0.684 & 0.2 & 0.78 & 0.813 & 0.31 & 0.0242 & 0.65 & 0.008 & 0.016 & -0.552 & 33.6 & 0.51 & 1.12 & -12.3 & 0.3 & 8.6 & 9.7 \\
C15 & 108.6 & 475.9 & 0.684 & 0.2 & 0.78 & 0.777 & 0.33 & 0.0240 & 0.36 & 0.008 & 0.016 & -0.530 & 32.3 & 0.49 & 1.15 & -6.3 & -0.3 & 4.4 & 16.3 \\
C16 & 110.8 & 8.6 & 0.684 & 0.2 & 0.78 & 0.833 & 0.19 & 0.0245 & 0.31 & 0.007 & 0.017 & -0.562 & 34.0 & 0.52 & 1.13 & -41.8 & 1.5 & 9.8 & -2.4 \\
C17 & 173.8 & 25.5 & 0.684 & 0.2 & 0.78 & 0.812 & 0.24 & 0.0239 & 0.66 & 0.007 & 0.017 & -0.550 & 33.9 & 0.51 & 1.10 & -42.0 & -0.8 & 9.5 & -2.3 \\
C18 & 317.2 & 64.2 & 0.684 & 0.2 & 0.78 & 0.819 & 0.17 & 0.0241 & 0.31 & 0.007 & 0.017 & -0.554 & 34.0 & 0.51 & 1.12 & -36.5 & -0.2 & 9.9 & -0.9 \\
C19 & 436.6 & 94.9 & 0.684 & 0.2 & 0.78 & 0.825 & 0.16 & 0.0241 & 0.33 & 0.007 & 0.017 & -0.558 & 34.3 & 0.51 & 1.15 & -32.9 & -0.2 & 10.6 & -0.2 \\
C20 & 543.1 & 126.9 & 0.684 & 0.2 & 0.78 & 0.820 & 0.16 & 0.0241 & 0.30 & 0.007 & 0.017 & -0.556 & 34.1 & 0.51 & 1.13 & -31.2 & -0.2 & 10.0 & 1.3 \\
C21 & 108.6 & 39.8 & 0.684 & 0.2 & 0.6 & 0.869 & 0.10 & 0.0253 & 0.34 & 0.007 & 0.018 & -0.583 & 34.3 & 0.53 & 1.15 & -38.1 & 4.9 & 10.8 & -9.7 \\
C22 & 108.6 & 39.8 & 0.684 & 0.2 & 0.4 & 0.884 & 0.31 & 0.0260 & 0.52 & 0.007 & 0.019 & -0.591 & 34.0 & 0.54 & 1.15 & -33.7 & 7.8 & 9.8 & -4.4 \\
C23 & 108.6 & 39.8 & 0.684 & 0.3 & 0.78 & 0.741 & 0.34 & 0.0240 & 0.32 & 0.008 & 0.016 & -0.508 & 30.9 & 0.47 & 1.08 & -32.1 & -0.7 & -0.1 & 6.1 \\
C24 & 108.6 & 39.8 & 0.684 & 0.4 & 0.78 & 0.713 & 0.16 & 0.0243 & 0.35 & 0.008 & 0.016 & -0.490 & 29.3 & 0.45 & 1.05 & -9.3 & 0.7 & -5.3 & 12.8 \\
C25 & 108.6 & 39.8 & 0.181 & 0.2 & 0.78 & 0.808 & 0.16 & 0.0243 & 0.37 & 0.008 & 0.017 & -0.548 & 33.2 & 0.50 & 1.14 & -16.6 & 0.9 & 7.2 & -4.4 \\
C26 & 108.6 & 39.8 & 0.363 & 0.2 & 0.78 & 0.852 & 0.18 & 0.0251 & 0.37 & 0.007 & 0.018 & -0.574 & 34.0 & 0.52 & 1.16 & -28.5 & 3.9 & 9.8 & -4.6 \\

\end{tabular}
}
\caption{The complete set of DNS simulations. 
The control parameters are shown in the left part of the table.
The middle part reports the aerodynamic coefficients ($C_L$ for lift, $C_D$ for drag and $C_M$ for moment) with the percentage relative standard deviation $\sigma$, the friction and pressure components of drag, $C_{D,f}$ and $C_{D,p}$, and aerodynamic efficiency $E$, shock position $x_s$ and shock intensity $I$.
The rightmost part shows the performance indicators discussed in the text.
The relative change of an indicator $\Phi$ is defined as the percentage increase with respect to the uncontrolled case (denoted with the `0' subscript), {\em i.e.} $\Delta \Phi = 100\left(\Phi/\Phi_0-1\right)$; in case R10, variations are computed against RREF.}
\label{tab:parameters}

\end{table}

\section{Aerodynamic performance}
\label{sec:aerodynamics}

In this section, we consider the effectiveness of StTW in terms of modification of the force and momentum coefficients of the airfoil, which are connected to changes in the distribution of the wall stresses along the chord. 
Last, the energetic performance of StTW is discussed. 

When discussing changes induced by the forcing, the variations, indicated with the symbol $\Delta$, are consistently defined as percentage increase with respect to the reference case.

\subsection{Force and moment coefficients}
\label{sec:forces}

\begin{figure}
\centering
\includegraphics{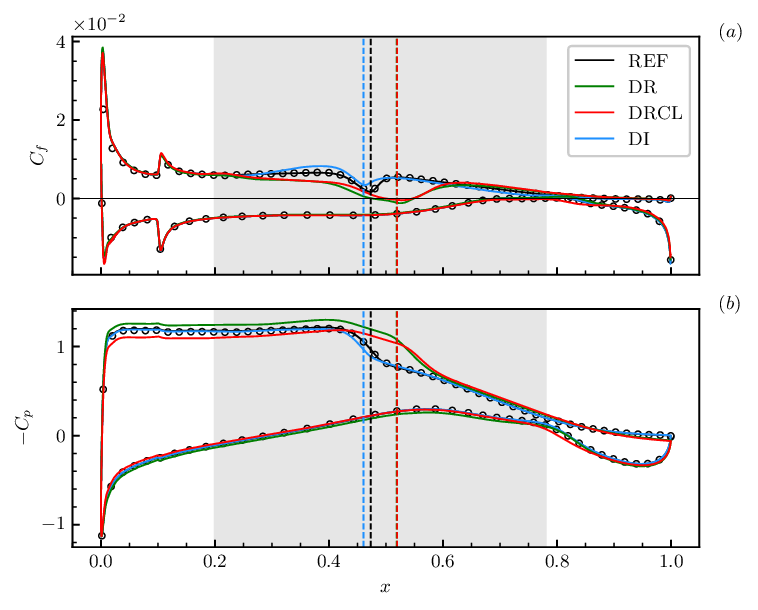}
\caption{Distributions of friction coefficient ($C_f$, panel $(a)$) and pressure coefficient ($C_p$, panel $(b)$) along the airfoil. 
Dots are from the reference simulation by \cite{quadrio-etal-2022}, and overlap with the present REF case.
The vertical lines mark the shock position for each case (note that DR and DRCL overlap); the grey area indicates the active region (on the suction side only).}
\label{fig:cfcp}
\end{figure}

Aerodynamic forces and moments derive from the integrated distributions of the projected wall friction and pressure, whose coefficients are defined as 
\begin{equation}
  C_f(x) = \frac{2 \tau_w(x)}{\rho_\infty U_\infty^2}, \qquad \qquad
  C_p(x) = \frac{2\left[p_w(x) - p_\infty\right]}{\rho_\infty U_\infty^2},
\end{equation}
where $\tau_w$ and $p_w$ are the mean wall-shear stress and the mean wall pressure, respectively, obtained after averaging in time and along the spanwise direction.

The mean aerodynamic force $\vect{F}$ is
\begin{equation}
\vect{F} = \oint_S \left[\tau_w \vect{\hat{t}} - p_w \vect{\hat{n}} \right] dS = 
\frac{1}{2} \rho_\infty U_\infty^2 \oint_S
\left[C_f \vect{\hat{t}} - C_p \vect{\hat{n}}\right] dS
\end{equation}
where $\vect{\hat{t}}$ and $\vect{\hat{n}}$ are the local unit vectors respectively tangent and normal to the wall, and $S$ is the wing surface.
$\vect{F}$ is conventionally decomposed into lift $L$ and drag $D$ components, respectively 
normal and parallel to the freestream velocity. 
The aerodynamic pitching moment $M_{TE}$ with respect to the trailing edge is determined as
\begin{equation}
M_{TE} = \vect{\hat{z}} \vect{\cdot} \oint_S \vect{r} \times \left[\tau_w \vect{\hat{t}} - p_w \vect{\hat{n}} \right] dS
\end{equation}
where $\vect{\hat{z}}$ is the unit vector in the spanwise direction and $\vect{r}$ is the distance vector from the trailing edge. The values of the force and momentum coefficients $C_L$, $C_D$ and $C_M$ are reported in table \ref{tab:parameters}.

Figure \ref{fig:cfcp}{\em a} shows the distribution of the friction coefficient along the chord. 
The reference case (black line) overlaps with the nominally identical one computed by \cite{quadrio-etal-2022}, with minor discrepancies near the leading edge due to differences in the numerical procedure adopted.
The plot highlights some important flow features, as the local peak of $C_f$ at $x=0.1$ due to tripping, the local reduction under the shock wave (whose position is indicated with the vertical dashed line), and the near zero values at the trailing edge on the suction side.
To analyze the effects of wall actuation on friction, the reference case is compared with cases DR, DRCL and DI. 
Direct effects of control are observed on the suction side only, where it is applied.
In terms of friction, the effects of StTW are as expected: after the forcing is applied (i.e. for $x>0.2$), friction decreases (or increases, for DI), becoming mildly negative past the shock for cases DR and DRCL.
Simulation DRCL with modified incidence is very similar to DR on either side of the airfoil, with exception of the region right upstream of the shock, where friction is slightly higher.
The shock moves downstream in the drag-reducing cases, whereas it moves upstream for DI.
Panel $(b)$ of figure \ref{fig:cfcp} shows the distribution of the pressure coefficient.
In the REF case, the sudden flow expansion at the leading edge is followed by a plateau where, by design, the airfoil achieves nearly zero pressure gradient. Further downstream, the mildly negative slope of $C_p$ reveals the presence of a shock. 
The wall pressure distributions for DR and DRCL do not coincide as a consequence of the different incidence, while the position of the shock across the two cases is unchanged.
Case DRCL provides the same lift as the reference case, although its wall pressure remains higher (hence with a lower $C_p$) than the reference simulation in the first 40\% of the chord: StTW displace the shock downstream, and the supersonic low-pressure region is wider.

\begin{figure}
\includegraphics{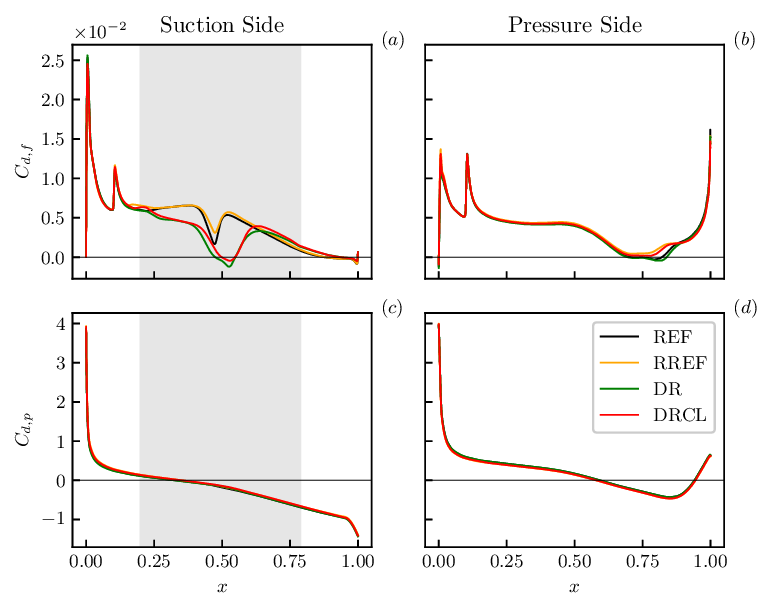}
\caption{Local friction ($C_{d,f}$, upper panels) and pressure ($C_{d,p}$, lower panels) contributions to $C_D$ on the suction side (left panels) and pressure side (right panels) of the airfoil. The grey area highlights the active region (on the suction side only).}
\label{fig:streamwise_Cd}
\end{figure}


Figure~\ref{fig:streamwise_Cd} shows the local friction and pressure contributions to the drag coefficient ($C_{d,f}$ and $C_{d,p}$, respectively, whose integrals are presented in table \ref{tab:parameters} as $C_{D,f}$ and $C_{D,p}$).
The distribution of $C_{d,f}$ is very similar to the friction coefficient (depicted in figure \ref{fig:cfcp}) on the suction side (panel $(a)$), as in this region the wall is rather flat and aligned to the free stream. 
The effect of wall actuation on the pressure side, shown in panel $(b)$, is instead rather small, with minor differences close to the trailing edge associated with different values of the angle of attack.
As for $C_{d,p}$ (bottom panels), all the distributions essentially overlap.
However, one should bear in mind the different magnitude of the contributions of pressure and friction to the overall drag, which renders pressure changes significant.

\begin{figure}
\includegraphics{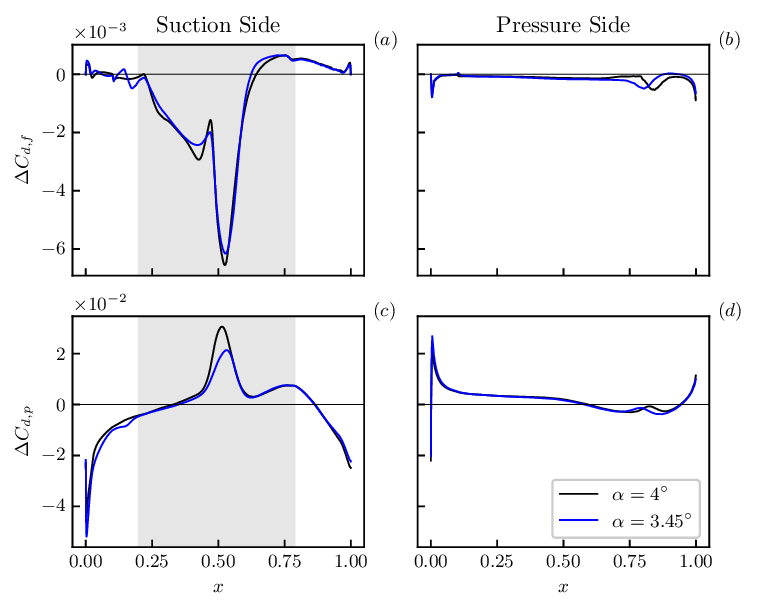}
\caption{Local difference of frictional ($\Delta C_{d,f}$, panels {\it a,b}) and pressure ($\Delta C_{d,p}$, panels {\it c,d}) contributions to the drag coefficient on the suction and pressure sides of the airfoil. The black lines express the differences between DR and REF cases, whereas the blue lines are obtained comparing cases at lower incidence DRCL and RREF.}
\label{fig:streamwise_DeltaCd}
\end{figure}

This is emphasized in figure~\ref{fig:streamwise_DeltaCd}, where differences between the controlled and uncontrolled cases at the two incidence angles of cases DR and DRCL are shown. 
On the suction side (panels $(a)$ and $(c)$), the frictional contribution only shows small variations, whereas the pressure contribution exhibits significant changes close to the shock, around the mid-chord. 
In that region, control yields positive variations of $C_{d,p}$ and contributes to drag increase. 
This positive contribution is smaller at the lower incidence $\alpha=3.45^\circ$, as the shock is weaker (see table~\ref{tab:parameters}).
Panels $(b)$ and $(d)$ show that changes on the pressure side are instead smaller, and do not depend on incidence.


Having isolated the effects of the reduced incidence angle from those directly ascribed to StTW, some conclusions can be drawn: 
(i) StTW always reduce friction drag, except for those combinations of control parameters for which drag increase is expected from channel flow information;
(ii) StTW tend to increase pressure drag, as a result of shock strengthening;
(iii) a smaller angle of attack reduces the pressure contribution to the drag force, at least within the range of values considered here, with a less significant effect on the friction contribution;
(iv) although the friction and pressure contributions to drag have different magnitudes, the relative control-induced variations are comparable.

\subsection{Effects of the control parameters}
\label{sec:parameters}
 
\begin{figure}
\centering
\includegraphics{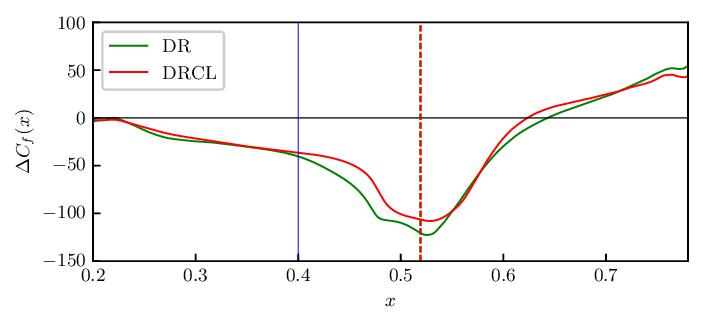}
\caption{Relative local change of the friction coefficient ($\DCf(x)$) on the portion of the suction side where control is active, for cases DR and DRCL compared to case REF. The vertical blue line at $x=0.4$ shows where frictional drag reduction is evaluated, whereas the two (nearly coincident) vertical dashed lines display the positions of the shock.}
\label{fig:rcf_x}
\end{figure}

The performance of StTW as a function of the control parameters is well understood in the canonical plane channel flow, where drag is entirely due to friction, and the mean friction is constant along the channel; comprehensive information exists which maps the control parameters into drag reduction figures.
Furthermore, even if compressibility is taken into account, the reduction of the skin-friction coefficient (and the net power gain) is essentially unchanged as long as the framework remains that of the plane channel flow, as demonstrated by \cite{gattere-etal-2024}.
hence, at least locally, the effects of StTW on the skin-friction over the transonic wing could be derived directly from the canonical channel flow.
However, the same does not apply on a global level, since the mean flow properties vary strongly along the streamwise direction due to wall curvature, adverse pressure gradient and the shock wave.

One can define the relative local reduction of the friction coefficient as
\begin{equation}
\DCf(x) = 100 \frac{C_f(x)-C_{f,0}(x)}{C_{f,0}(x)} ;
\end{equation}
this quantity varies because of changes of $C_f$ itself, and because of the presence of the gradual spatial evolution after $x_b$ of the control effect \citep{skote-2012}.
Figure~\ref{fig:rcf_x} plots $\DCf(x)$ for flow cases DR and DRCL, limited to the portion of the suction side where forcing is active. 
The variations of $C_f$ along the chord are significant, thus making it difficult to distill the function $\DCf(x)$ into a single number to compare with the case of the channel flow. 
However, one can identify a portion of the curve (for example $x \le 0.4$) over which changes in friction can be attributed mainly to wall actuation.
Hence, we consider the value of $\DCf$ at $x=0.4$ as a surrogate measure of frictional drag reduction, and define the viscous units with the friction velocity measured at $x=0.4$.
A detailed analysis of how this choice affects the comparison with the case of channel flow is presented in Appendix~\ref{app:friction}, together with law-ot-the-wall plots of the mean streamwise velocity profiles extracted at $x=0.4$.

\begin{figure}  
\centering
\includegraphics{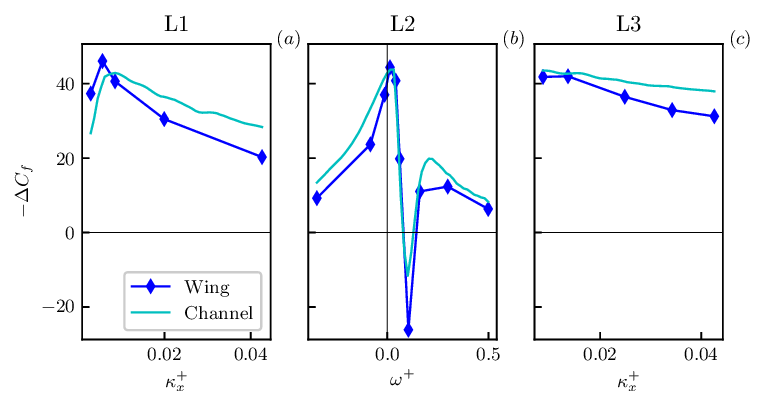}
\caption{Comparison of local reduction of friction coefficient ($- \DCf$) for flow cases C1 -- C20, across lines L1, L2 and L3 in figure \ref{fig:paramspace} and incompressible channel flow at $\Rey_\tau = 200$~\citep{gatti-quadrio-2016}.}
\label{fig:ChannelvsTragfoil}
\end{figure}

In figure~\ref{fig:ChannelvsTragfoil} we compare the values of friction drag reduction $-\DCf$ at $x=0.4$ for the simulations C1 -- C20 at constant forcing intensity $A$ with data for the incompressible channel flow at $\Rey_\tau = 200$~\citep{gatti-quadrio-2016}, properly interpolated to precisely match the control parameters of each case.
While the magnitudes of $\DCf$ should be regarded qualitatively for the reason discussed above, the trends are remarkably similar. 
In particular, in both cases the global maximum of drag reduction is found for slow, forward-travelling waves.

\begin{figure}
\centering
\includegraphics{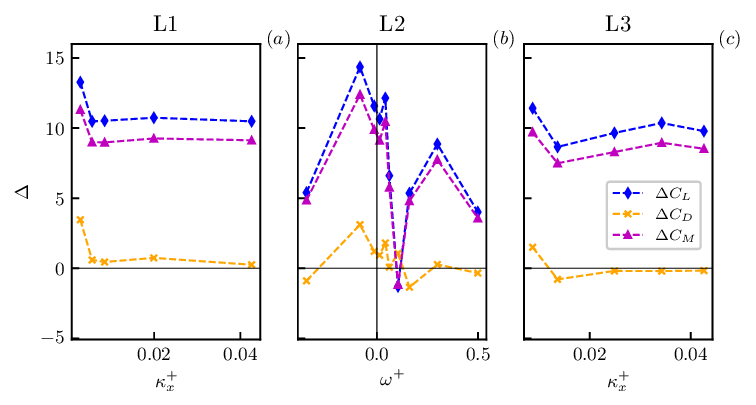}
\caption{Relative variations of lift ($\Delta C_L$), drag ($\Delta C_D$) and pitching moment ($\Delta C_M$) coefficients, for the cases C1--C20 across lines L1, L2 and L3 of figure \ref{fig:paramspace}.}
\label{fig:forces}
\end{figure}

Relative variations of the force and momentum coefficients of the airfoil 
for flow cases C1 -- C20 are reported in figure~\ref{fig:forces}.
Perhaps surprisingly, the drag coefficient is found to undergo little changes, with at most a reduction of approximately $1\%$, as a consequence of wall actuation; in addition, in some cases it even increases slightly. 
As expected, changes in the pitching moment, which is mainly controlled by pressure, are analogous to changes in $C_L$.

\begin{figure}
\centering
\includegraphics{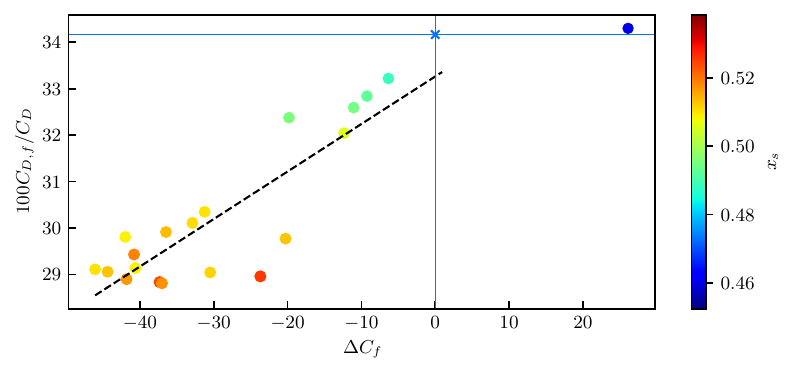}
\caption{Percentage friction contribution to drag ($100 C_{D,f}/C_D$), 
as a function of the friction drag changes ($\DCf$)  
for cases C1--C20 across lines L1, L2 and L3 of figure \ref{fig:paramspace}. The color of each data point encodes the shock position. 
The cross indicates the reference case, while the dashed line is a linear regression of the data for $\DCf < 0$.}
\label{fig:CDbudget}
\end{figure}

Figure~\ref{fig:CDbudget} displays the friction contribution to the total drag combined with the friction drag reduction.
StTW generally decrease the friction contribution to the overall drag, with the exception of the friction-drag-increasing case DI.
Whereas in the reference simulation friction is responsible for $34\%$ of the entire drag, this fraction reduces to $29\%$ in the presence of wall actuation. 
The minor changes observed in $C_D$ imply that the relationship between $\DCf$ and $C_{D,f}$ is close to linear.
The shock wave is located at $x_s=0.47$ in the reference case, and a connection between a large local friction reduction and a downstream displacement of the shock is clearly visible.
The further the shock is delayed towards the trailing edge, the stronger the expansion on the suction side of the airfoil.
Since the post-shock conditions to be matched are approximately the same (see $C_p$ in figure~\ref{fig:cfcp}), we can conclude that the shock intensity does depend on the shock position, as can be seen from the parameters listed in table~\ref{tab:parameters}.
Hence, a stronger shock leads to a larger pressure contribution to the aerodynamic coefficients.

\subsection{Net savings}

\begin{figure}
\centering
\includegraphics{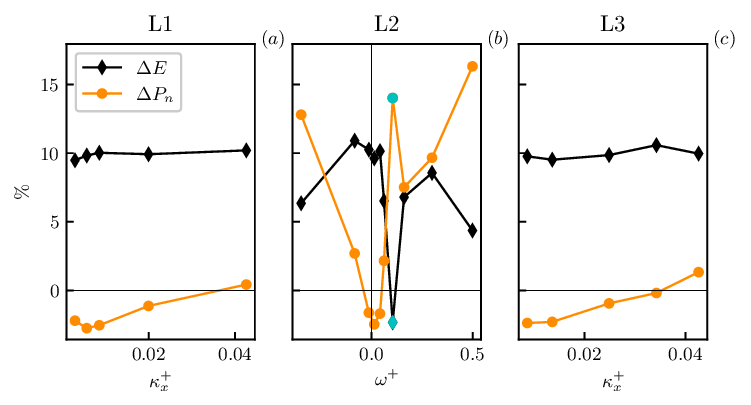}
\caption{Changes in aerodynamic efficiency ($\Delta E$) and net power ($\Delta P_n$) for cases C1 -- C20 across lines L1 (standing wave), L2 (fixed $\kappa_x$) and L3 (ridge of maximum drag reduction) of figure~\ref{fig:paramspace}. The cyan markers highlight case DI.}
\label{fig:net}
\end{figure}

As discussed in \S\ref{sec:introduction}, assessing the effectiveness of active flow control strategies requires to estimate the energy expenditure for the actuation.
In canonical flows, in which drag consists only of friction, one may want to exploit the positive effect of friction-reducing control to improve the balance between performance (e.g. the mass flow rate in channel flow, or the pressure losses through the duct) and the related cost, the pumping power.
However, infinite ways exist to modify such balance, the two extreme cases being maximization of performance for given cost, or minimization of cost for given performance. 
In a plane channel flow~\citep{frohnapfel-hasegawa-quadrio-2012} the two approaches are referred to as constant-pressure-gradient comparison (CPG, whereby the pumping force per unit volume is unchanged and the flow rate can change) and the constant-flow-rate comparison (CFR, whereby the flow rate is unchanged and the pumping force per unit volume can change).
A similar line of reasoning applies to the present case, 
but the definition of costs and benefits is less obvious. 
As we have seen, the main effect of wall actuation
is to increase lift; however, lift a poor indicator of performance. 
In fact, in cruise flight the wing is bound to always generate the same lift force for balancing the aircraft weight.

In analogy with the CFR comparison procedure, an increase of lift can be exploited to reduce the angle of attack, thus aiming at cost minimization for unchanged performance. 
Let $\Pi_0 = D_0 U_\infty$ be the nominal power required for flying the aircraft at 
constant speed $U_\infty$, with $D_0$ the drag force in the uncontrolled configuration.
One can define the net rate of change of cruise flight power requirement ($\DP$) as the ratio between the reduction of flight power (incremented by the control power), and the reference power:

\begin{equation}
\DP = \frac{\left(\Pi + \Pi_c\right) - \Pi_0}{\Pi_0}, 
\label{eq:NPS}
\end{equation}
where $\Pi = D \, U_\infty$ is the power required to fly the wing at the same speed $U_\infty$ 
in the controlled configuration.
The subscript $n$ in $\DP$ emphasizes that the power change is a net balance that accounts for the extra power $\Pi_c$ required by the active forcing.
At the numerator of equation~\eqref{eq:NPS}
the power reduction $-\left(\Pi - \Pi_0\right)$ is obviously linked to the drag 
reduction ($-\left(D - D_0\right)$), where the drag $D$ is the drag obtained
in a configuration yielding the same lift of the reference case.
Finally, the control power $\Pi_c$ is the power transferred from 
the StTW to the fluid, assuming unitary efficiency $\gamma=1$ of an ideal actuator, as alway done in numerical studies.
Values for $\gamma$ are typically not provided by experimental studies \citep[e.g.][]{bird-santer-morrison-2018, marusic-etal-2021, fumarola-santer-morrison-2024, knoop-etal-2024} where StTW were tested; these studies either overlook the net savings, or simply assume ideal efficiency. While \cite{gatti-etal-2015} mention a large value of $\gamma=0.7$ for their electro-mechanical actuator for the spanwise oscillating wall, the typical proof-of-principle experiment employing simple mechanical actuators has low efficiencies: for example, actuation power reported by \cite{auteri-etal-2010} for their mechanical implemention of StTW is 1528 milliWatts, against an ideal control power of 2.1 milliWatts ($\gamma=0.0014$).

The drag $D$ required to evaluate equation~\eqref{eq:NPS} is derived from the lift increase $\Delta L$ brought about by wall actuation via the following procedure:
(1) the lift curve $C_L(\alpha)$ of the uncontrolled wing is used to determine the reduced incidence angle required to recover the original lift when the control is applied (hence causing the variation $\Delta C_L$ available in table \ref{tab:parameters});
(2) the drag curve $C_D(\alpha)$ of the uncontrolled wing is used to interpolate drag at the reduced incidence (the presence of forcing is not accounted for yet);
(3) the drag coefficient at the reduced incidence is modified to account for the variation $\DCd$ (available in table \ref{tab:parameters}) caused by the direct action of StTW.

Points (1) and (2) above require the polar curve of the V2C airfoil, which is impractical to obtain via DNS, and has been computed with RANS instead. 
In particular, the open-source compressible RANS solver SU2 v.7.5.1 is used, with differential operators at second-order accuracy, and free-stream conditions matching those of the DNS.
The central Jameson--Schmidt--Turkel scheme \citep{jameson-schmidt-turkel-1981} has been employed for the convection terms, with the default Venkatakrishnan slope limiter \citep{venkatakrishnan-1995} employed to preserve monotonicity near shocks.
The weighted least squares method is employed for discretising the gradient operator involved in the viscous terms.
Turbulence is modeled through the wall-resolved Spalart--Allmaras model \citep{spalart-allmaras-1992}.
The grid is an unstructured, two-dimensional circular mesh with radius of $100$ chords, consisting of 429100 volume elements.
It features an orthogonal prism layer refinement at the airfoil surface, which guarantees element orthogonality and a wall resolution of approximately one viscous unit just ahead of the shock.
In order to avoid additional uncertainty, no transition modelling has been employed.
The simulations are run until the relative change of drag coefficient is smaller than $10^{-8}$.
The pseudo timestepping required to reach convergence of the steady simulation is performed with implicit Euler.

Results of the procedure in terms of $\Delta P_n$ are presented in figure~\ref{fig:net}: we recall that under the convention employed in this paper, positive $\Delta P_n$ implies more power required in the controlled case, hence power loss.
Figure \ref{fig:net} shows that changes of aerodynamic efficiency are generally positive with $\Delta E$ typically increasing by $10\%$ on average and up to 30\%. An exception is case DI, highlighted by the cyan markers, which yields drag increase and consequently reduced efficiency.
Against the improved efficiency, the required power does not always decrease, being basically unchanged in a few cases and increased in the remaining ones.
This is an expected result that parallels the incompressible channel flow, where the power balance at this rather large forcing amplitude is unfavourable.
As in the channel, the net balance is found to be best for slow, forward-travelling waves \citep{quadrio-ricco-viotti-2009}.

Extent and position of the active region are two control parameters with no counterpart in the homogeneous channel flow.
Simulations C21 -- C24 have been performed to quantify how the baseline values of $x_b=0.2$ and $x_e=0.78$ considered by~\cite{quadrio-etal-2022} affect the results. 
Specifically, in flow cases C21 and C22 the end point of the actuated region is moved upstream for fixed $x_b$, whereas in cases C23 and C24 the origin is shifted downstream for a fixed $x_e$. 
In all cases the active area is reduced, yet the results listed in 
table~\ref{tab:parameters} show that $\DCf$ is minimally altered when $x_e$ 
is shifted upstream; on the contrary, moving $x_b$ downstream  
adversely affects $\DCf$, $x_b=0.4$ being the worst case.
The best choice for increasing the aerodynamic efficiency seems thus to be
$x_b=0.2$ and $x_e=0.6$, yielding $\DP = -9.7\%$, which is the best 
value obtained among all the DNS performed at $\alpha=4^\circ$.
\begin{figure}
\centering
\includegraphics{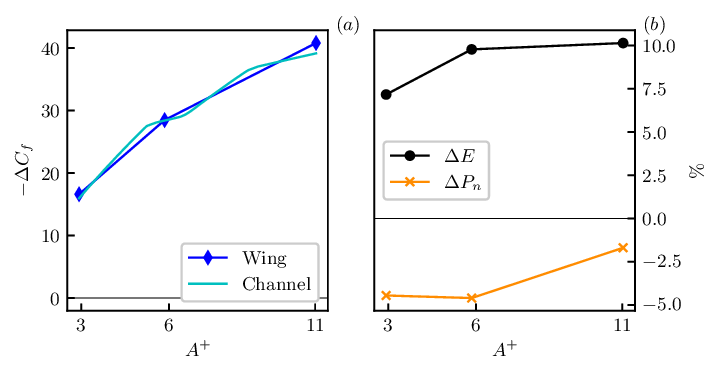}
\caption{Performance metrics for flow cases with different forcing amplitude $A^+$: C25 ($A^+ \approx 3$), C26 ($A^+ \approx 6$) and DR ($A^+ \approx 11$). Panel $(a)$: changes in $C_f$, compared with incompressible channel flow data; panel $(b)$: changes in the aerodynamic efficiency and net savings.}
\label{fig:net2}
\end{figure}
 
It is well known that the energetic efficiency of StTW varies with the forcing amplitude $A$ in the case of flow over plane walls.
This dependence is here verified in flow cases C25 and C26, which 
replicate case DR with lower forcing amplitudes.
The results are presented in figure~\ref{fig:net2}, in which panel $(a)$ shows the resulting changes of $\DCf$ compared with the incompressible channel flow data, interpolated from the database by \cite{gatti-quadrio-2016}, while panel $(b)$ displays $\DE$ and $\DP$.
Just as for incompressible channels, the friction drag reduction decreases for smaller forcing amplitudes.
The agreement, both in qualitative and quantitative terms, between the curves of wing and plane channel flow is remarkable.
Interestingly, the improvement of the aerodynamic efficiency (panel $(b)$)
does not follow $\DCf$, and halving the control intensity from $A^+ \approx 11$ 
to $A^+ \approx 6$ does not decrease $\Delta E$. 
We find a minumum value of $\DP$ for the case C27, corresponding to $A^+=6$. 
This is a significant analogy with the incompressible channel flow, in which maximum efficiency is obtained for $A^+ \approx 5$ \citep{quadrio-ricco-viotti-2009}, as the power cost grows quadratically with $A$.

Finally, it is important to stress that in this specific application the true benefits of StTW hinge upon the increase of aerodynamic efficiency, and become more significant when the entire aircraft is considered instead of the wing alone.
Therefore, for case DR, we replicate the line of reasoning followed by \cite{quadrio-etal-2022}, and extrapolate the potential saving to a complete aircraft.
Specifically, we consider the same wing-body configuration, i.e. the `DLR-F6' \citep{laflin-etal-2005}, at the reference uncontrolled conditions of $M=0.75$, $Re_\infty=3 \times 10^6$ and $\alpha_0=0.52^\circ$. Aircraft information is extracted from the dataset available online at \url{https://aiaa-dpw.larc.nasa.gov/Workshop2/DPW_forces_WB_375}.

The simplifying hypotheses adopted are the following: 
\begin{enumerate}
\item the wing-generated lift is the only contributor to the overall lift of the aircraft: the fuselage and the tail do not contribute (hence, no additional lift sources are considered);
\item the non-lift-induced drag (i.e. the drag that is computed via this procedure) corresponds to one-third of the overall drag; 
\item the aerodynamic coefficients and the control effects are constant along the wingspan, and do not change with $Re_\infty$ and $M_\infty$, so that results obtained for the wing section are extended to the whole wing without finite-wing corrections;
\item the slopes of the curves $C_L-\alpha$ and $C_D-\alpha$ are not affected by the control;
\item the actuated surface is one-fourth of the wing surface, and one-twelfth of the whole aircraft surface.
\end{enumerate}

Under these assumptions, one can estimate how StTW as in case DR would improve the drag coefficient of the entire aircraft.
The procedure adopted, that resembles the one described above to obtain $\DP$, has already been presented in \cite{quadrio-etal-2022}, and therefore is not reported here in detail.
The smaller angle of attack in cruise flight, made possible by StTW, translates into a decrease of the aircraft drag coefficient of $\Delta C_D=-8\%$.
This reduction is then augmented by the effect of the StTW at the reduced angle of attack (see table \ref{tab:parameters}), which yields the additional benefit of $\Delta C_D = -4.4\%$.
From the energetic viewpoint, the theoretical $\gamma=1$ control power $\Pi_c$ required by the actuation at unitary efficiency is $0.4\%$ of the overall flight power, yielding a nominal net gain of 12\%.
While a suitable actuator for StTW is missing, it may be interesting to mention that this combination of net savings and required power translates into the minimum value of efficiency required by an actuator to provide a positive power gain. 
The mimimum efficiency is simply the inverse of the maximum gain discussed by \cite{kasagi-hasegawa-fukagata-2009}. 
If the requirement for net savings is reformulated via equation \eqref{eq:NPS} for the complete aircraft and for non-unitary control efficiency as
\begin{equation}
\Delta P_n = \frac{\left(\Pi + \Pi_c / \gamma \right) - \Pi_0}{\Pi_0} \le 0
\label{eq:etamin}
\end{equation}
the control performance is such that an actuator efficiency as low as $\gamma = 0.045$ is enough to yield a net power gain.


\section{Flow organization}
\label{sec:flowphys}

After the description of the effect of StTW in global terms through the aerodynamic coefficients, the focus is shifted here on the interaction between the forcing (or, more precisely, the reduced levels of friction brought about by it) and the turbulent boundary layer with the shock wave.

\subsection{Boundary layer} 
\label{sec:evolution}
\begin{figure}
\centering
\includegraphics[width=\textwidth]{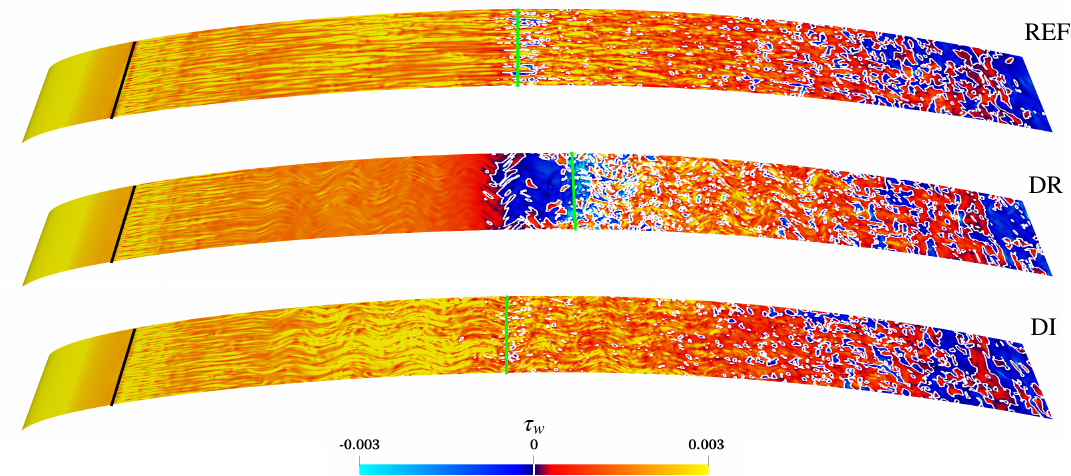}
\caption{Instantaneous wall-shear stress on the suction side of the airfoil. The black lines indicate the position of the tripping, and the green lines denote the position of the shock in the three cases; the white contours correspond to $\tau_w = 0$.}
\label{fig:tau_flow}
\end{figure}
Figure~\ref{fig:tau_flow} shows a comparative view of the instantaneous 
wall shear stress ($\tau_w$) on the suction side of the wing 
for the REF, DR and DI flow cases.
Close to the leading edge, the flow is laminar up to the tripping location, where streaks of high- and low-speed fluid, whose footprints are visible at the wall, begin to form due to the imposed transition.
In the reference case, turbulent structures are convected downstream towards the shock, 
where the strong adverse pressure gradient slows the flow and changes their shape.
The wall-shear stress remains positive on average around the shock, 
while showing instantaneous, locally negative spots. 
In case DR, the streaks are similar to the reference case right past
the tripping point, but the forcing affects their dowstream development past $x_b=0.2$.
Further downstream, where the influence of the shock is stronger, mild flow reversal 
develops upstream of the shock position, as made evident by the large region with negative wall shear.
The observed flow reversal however is likely connected with the limited 
Reynolds number of the present configuration.
In flow case DI, in which wall actuation is ineffective and even yields local increase of $\tau_w$, the near-wall cycle is strengthened, yielding turbulence intensification 
and higher friction. Here, the streamwise modulation of the flow induced by 
wall actuation is clearly visible, both upstream and downstream of the shock. 
According to \citet{quadrio-ricco-viotti-2009}, this is an indicator that spanwise forcing is operating far from the optimal conditions.

\begin{figure}
\centering
\includegraphics{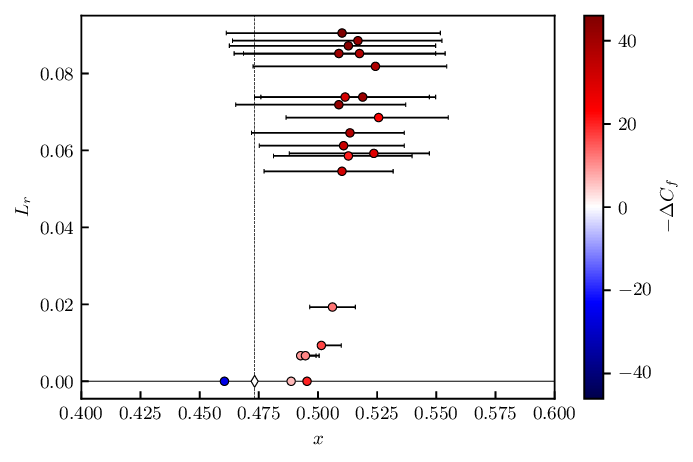}
\caption{Shock position (dots) and extent of the reverse-flow region (horizontal bars). The vertical dashed line marks the shock position in the reference case. The dots are coloured according to the skin-friction reduction rate $-\DCf$. The white diamond corresponds to the unmanipulated flow.}
\label{fig:bubble-length}
\end{figure}

The reversed flow region around the shock is further explored in figure \ref{fig:bubble-length}, where its length and position, color-coded with $-\DCf$, are plotted against the position of the shock wave $x_s$ for cases with constant control extension (i.e. cases C1--C20 and C25--C26).
Figure \ref{fig:bubble-length} highlights a dependence between the length of the separated region and the position of the shock wave, further influenced by the skin-friction reduction rate; note that this relationship is quite robust, as the plot includes also results with different control amplitude (cases C25 -- C26).
In fact, we observe that the reversed flow area consistently includes the shock, and all cases with large $\DCf$ feature long separation bubbles.
The longest reversed flow region (about 9\% of the chord) is found in case C2, which also achieves the largest friction drag reduction ($\DCf=-46\%$).
Conversely, case DI (drag-increase) and case C15 (which has the smallest positive $-\DCf$) exhibit no flow separation and the shock resides further downstream, similarly to the unactuated case.

\begin{figure}
\centering
\includegraphics{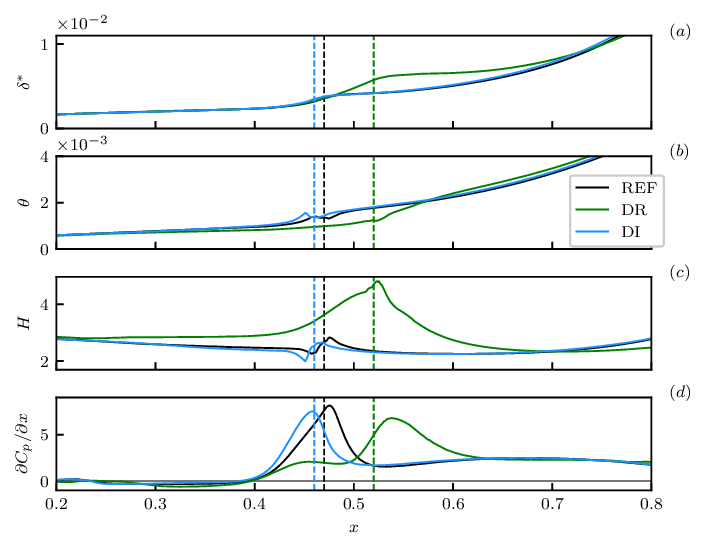}
\caption{Distribution of integral boundary layer parameters
and pressure gradient across the actuated region on the suction side
for the REF, DR and DI flow cases. From top to bottom: 
displacement thickness ($\delta^\ast$), momentum thickness ($\theta$), 
shape factor ($H$), and longitudinal component of the pressure gradient 
($\partial{C_p}/\partial{x}$).}
\label{fig:thicknesses}
\end{figure}

The compressible boundary layer thicknesses $\delta^*$, $\theta$ and shape factor $H$ are are shown in figure \ref{fig:thicknesses}. 
It is well known that the local wall-shear stress is linked to streamwise variations of the boundary layer thicknesses $\delta^\ast$ and $\theta$ via the von K\'arm\'an integral equation.
They are computed as in \cite{xu-ricco-duan-2023}:
\begin{equation}
	\delta^\ast=\int_0^\infty \left(1-\frac{\overline{\rho}~\widetilde{u}}{\rho_e U_e}\right) d\eta
	\qquad
	\theta = \int_0^\infty\left(1-\frac{\widetilde{u}}{U_e}\right) \frac{\overline{\rho}~\widetilde{u}}{\rho_e U_e}d\eta
	\qquad
	H=\delta^\ast/\theta
	\label{eq:blheights}
\end{equation}
where $\overline{\cdot}$ and $\widetilde{\left(\cdot\right)}$ denote the temporal and Favre-averaging operators, and the subscript `e' denotes a quantity at the edge of the boundary layer.
The edge of the boundary layer is identified with the wall-normal position at which the tangential velocity reaches 99\% of the maximum local tangential velocity.


As long as the wall curvature is limited, a reduction in the wall-shear stress should be associated with a reduction in the growth rate of the momentum thickness.
This is indeed observed in the curve corresponding to case DR in panel $(b)$ of figure~\ref{fig:thicknesses}.
Moreover, the pre-shock, adverse pressure gradient region in case DR is wider, the shock moves downstream significantly, and the deceleration required to bring the flow back to the trailing edge conditions is stronger:
the larger velocity defect is reflected in a larger displacement thickness,
which is indeed seen to grow significantly at $x \simeq 0.5$. 
The combination of these two effects yields a higher shape factor in flow case DR, which 
indicates that the boundary layer is more prone to flow reversal.
The drag-increasing case DI differs in that
near the shock, both $\delta^*$ and $\theta$ (panels $(a)$ and $(b)$, respectively) slightly grow due to the adverse pressure gradient, hence the shape factor, panel $(c)$, does not change significantly. 
The streamwise pressure gradient is shown in panel $(d)$ of figure~\ref{fig:thicknesses}. 
For the reference and drag-increasing cases, the peak of $\partial{C_p}/\partial{x}$ 
corresponds to the shock position, whereas for flow case DR the peak lies distinctly downstream 
of the shock, with approximately the same magnitude.
A barely discernible secondary peak is also found upstream of the shock 
($x \approx 0.47$), the two peaks approximately marking the start and the end of 
the flow reversal.

\begin{figure}
\centering
\includegraphics{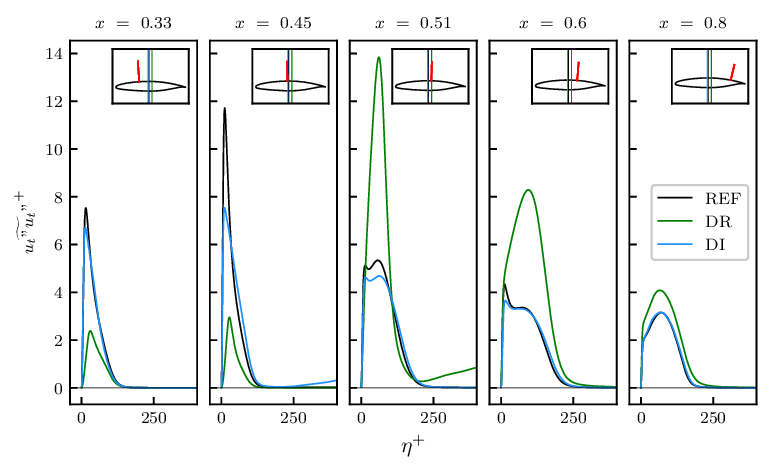}
\caption{Profiles of tangential turbulent stress ($\widetilde{u_t''u_t''}$) along the wall-normal coordinate $\eta$, at five positions along the chord (whose streamwise coordinate is reported on top of each panel) for the REF, DR and DI flow cases. In the insets, the red lines denote the position along the chord and the orientation of the normal, while the vertical coloured lines show the positions of the shocks.}
\label{fig:R11}
\end{figure}

Last, we have inspected the Reynolds stress tensor to obtain further 
insight into the turbulence activity within the boundary layer. 
The components of the Favre-averaged Reynolds stress tensor 
($R_{ij}=\widetilde{u_i''u_j''}$, where $\left(\cdot\right) ''$ denote the Favre fluctuating component) have been determined by locally decomposing the velocity field into tangential ($t$), normal ($n$) and spanwise ($z$) velocity components.
Figure~\ref{fig:R11} shows the profiles of the tangential stress $R_{tt}$ scaled in reference viscous units at
five stations along the suction side.
In the reference case, the stress is maximum upstream of the shock, 
($x \approx 0.45$, second panel), the amplification being
related to the formation of a mixing layer~\citep{pirozzoli-etal-2010}.
\cite{fang-etal-2020} have shown that in the decelerating region upstream 
of the foot of the shock, the adverse pressure gradient decelerates the boundary 
layer, thus increasing the production of turbulence kinetic energy via the longitudinal 
production term $-\widetilde{u_t''u_t''}\partial{\widetilde{u}}/\partial{\xi}$. 
Hence, a positive feedback loop is established for the turbulent kinetic energy, 
and in particular for $\widetilde{u_t''u_t''}$, until the 
flow stops decelerating downstream of the shock.
For simulation DR, the local maximum of $R_{tt}$ takes place further downstream, 
at $x=0.51$ (third panel). Upstream of the shock, case DR exhibits 
nearly halved fluctuations, whose peak is shifted away from the wall 
(from $\eta^+ \approx 10$ of the reference case to $\eta^+ \approx 30$). 
Flow case DI instead exhibits the opposite behaviour, 
with slight reduction of the peak, which also moves closer to the wall. 
Differences are striking at $x=0.51$, which is upstream of the shock 
in case DR, but downstream of it in cases REF and DI.
Looking at the two rightmost panels, the growth of the
boundary layer is reflected in the larger wall distance of the peaks, 
which occur at $\eta^+ \approx 90$. The monotonic decrease of 
$\widetilde{u_t''u_t''}$ confirms the gradual spreading of the mixing 
layer \citep{pirozzoli-etal-2010}.

\subsection{Transient analysis}
\label{sec:transient}

\begin{figure}
\vspace{5pt}
\centering
\includegraphics[width=\textwidth]{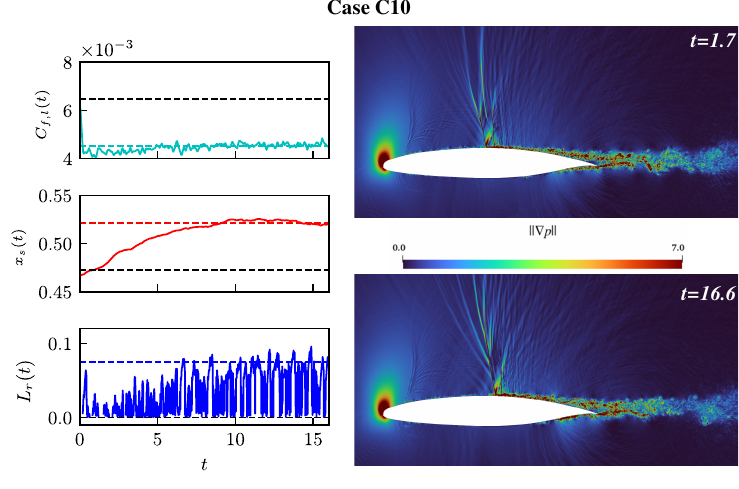}
\label{fig:bubble_bestDR}
\vspace{8pt}
\includegraphics[width=\textwidth]{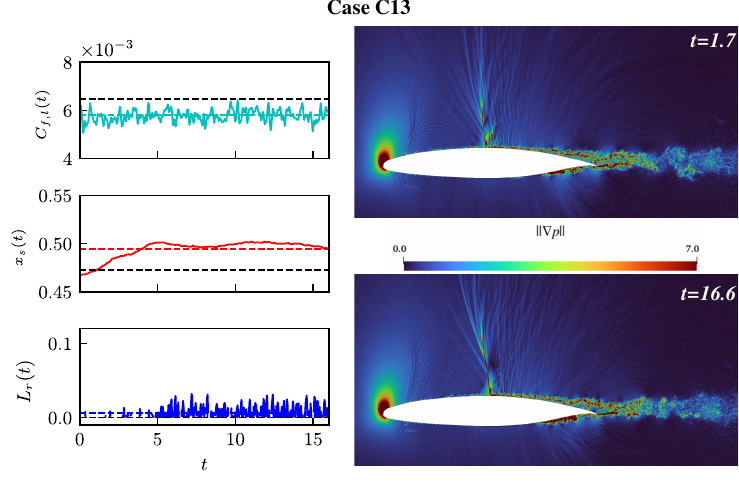}
\label{fig:bubble_noshock2}
\caption{
Transient of flow cases C10 (top) and C13 (bottom).
On the left: time histories of the spanwise-averaged mean friction coefficient between $x=0.3$ and $x=0.4$ (${C_{f,l}(t)}$, top), the shock position ($x_s(t)$, center) and the length of the recirculation bubble ($L_r(t)$, bottom), after sudden activation of StTW.
The horizontal lines show the mean asymptotic value (coloured) and the mean value of the REF flow case (black).
On the right: magnitude of the pressure gradient $\| \vect{\nabla} p\|$ on a $(x,y)$ plane, shortly after the control activation ($t = 1.7$), and at a later time ($t=16.6$).
}
\label{fig:transients}
\end{figure}

To further elucidate the relationship between StTW and flow reversal, we have analyzed the transient following the sudden application of StTW at $t=0$.
Specifically, flow samples separated in time by $\Delta t = 6.2 \times 10^{-3}$ have been collected over a time span of $16.6$ convective time units after wall actuation is turned on.
Figure~\ref{fig:transients} describes flow cases C10 (denoted as `DR' so far) and C13; the former has a large reversed flow region, whereas in the latter the region of reversal is significantly smaller. 
The left panels show the time history of spanwise-averaged quantities, additionally smoothed out by a running average over 6 time samples.
The quantities are, from top to bottom, the local suction-side friction coefficient ($C_{f,l}$), defined here as the average over the streamwise interval $0.3 \le x \le 0.4$, the shock position $x_s$ and the flow reversal length ($L_r$).
The right panels plot two snapshots of the magnitude of the instantaneous field of the pressure gradient, $\| \vect{\nabla} p \|$, plotted on a $(x,y)$ plane at $t=1.7$ and at $t=16.6$. 

In both simulations, after enabling control at $t=0$, the friction coefficient 
approaches its asymptotic value on a time scale that is short enough 
to be barely visible in the plots (top-left panels).
A non-equilibrium state then ensues, as the reduced friction causes the shock to move downstream.
This takes place on a much longer time scale, 
corresponding to about eight convective time units for flow case C10 and four time units for flow case C13. 
Flow reversal is present in flow case C10 since the earliest stages, 
intermittently at first, then growing as the shock moves downstream,
and becoming nearly stationary for $t\gtrsim 8.2$.
Flow case C13 initially features intermittent flow reversal, with relatively long time intervals with fully attached flow.
Approximately five time units are needed for flow reversal to fully establish. 
Since the shock is located more upstream in C13, the intensity and extent of the adverse pressure gradient region is less, hence the final length of the reversed flow region is significantly smaller than for case C10.

Flow reversal past the shock is only an aspect of transonic shock wave / turbulent boundary layer interaction, that was analyzed via DNS by \cite{pirozzoli-etal-2010}.
Turbulent boundary layers hit by a shock tend to thicken upstream of the interaction owing to the strong pressure rise caused by the shock. 
This creates compression waves, and, if the Mach number is large enough, induces a separation, ultimately resulting in a $\lambda$-shock configuration~\citep{delery-marvin-1986}.
The right panels of figure~\ref{fig:transients} display two different shock patterns. 
For flow case C10, at the short time $t = 1.7$ the lower part of the shock tilts, generating compression waves travelling downstream towards the trailing edge. 
During this motion, they encounter and merge with pressure waves from the wake at the trailing edge, resulting into a secondary, smaller shock, 
which stabilizes quickly. Exchange of information via pressure waves 
takes place between the two shocks through both the supersonic and subsonic regions, 
until the system reaches a stable configuration around $t = 16$.
It is worth noting that the final shock configuration resembles the typical topology of neither a weak shock/boundary layer interaction (i.e., with a single shock), 
nor a strong one (in which a $\lambda$-shock is formed).
In flow case C13 we observe intensification of the shock 
with respect to its initial configuration, without the secondary shock.
The time evolution from $t=1.7$ onwards shows that pressure waves depart 
from the main shock and merge with those generated at the trailing edge, 
but they do vanish at $t \approx 8$. 

Therefore, we conclude that StTW affect first the turbulent boundary layer directly, 
on a very short time scale, by abating the friction coefficient locally. 
On a longer time scale, the reduced friction indirectly causes the shock to move towards the trailing edge; this induces global modifications 
of the wall pressure distribution and then of the entire boundary layer. 
Flow reversal, which seems to be more a consequence of the modification of the 
shock position rather than a direct effect of wall actuation, causes the development of a secondary supersonic region. Whenever the recirculating bubble is large enough, a second, 
weaker shock develops.

\section{Concluding discussion}
\label{sec:conclusions}

In this work, we have extended the study by 
\cite{quadrio-etal-2022}, describing the turbulent transonic flow around 
a three-dimensional wing slab where flow control made by StTW of spanwise wall forcing 
is employed to improve the aerodynamic performance of the wing.
Several direct numerical simulations have been carried out to explore 
the space of the control parameters, to compare the present results 
with the incompressible channel flow, and to assess the effectiveness 
of StTW in a typical aeronautical application.

Our findings confirm that StTW, besides friction, also affect the other contributor 
to aerodynamic drag, i.e. pressure. The main consequence of StTW is a varied position of the suction-side shock, which is shifted toward the trailing edge when the control parameters are properly selected to reduce friction.
As a result, lift, drag and moment coefficients of the wing are all modified 
by the control. Interestingly, while lift increases significantly, drag is not much affected by StTW, even though the forcing 
reduces friction locally as expected.
This is due to an increase in pressure drag, resulting from an intensification 
of the shock. In fact, changes induced by StTW to friction and pressure drag 
are comparable in magnitude.
By studying the local contributions to the changes of the drag coefficient, 
we have been able to decouple the direct effects of the forcing from those 
deriving from the reduced angle of attack required to obtain the same lift.
The aerodynamic efficiency of the airfoil can rise up to $11\%$, 
leading to a large reduction in the drag coefficient of the entire aircraft, together with a negligible actuation power.
A qualitative estimate has placed the potential net savings for the entire aircraft at about 12\%.
Such a large improvement of the aerodynamic performance implies that net savings could still be obtained with an actuator of efficiency as low as $0.045$, thanks to the large benefits and the comparatively small amount of required energy, owing to the limited actuation area.

One question that we tackled, at least qualitatively, 
is whether our understanding of the optimal parameters for StTW in the 
incompressible channel flow can be brought forward to the present, more complex application.
Similarities with channel flows have been found in terms of friction 
drag reduction: by varying the control parameters, the reduction of the wall-shear stress
evolves in a way that resembles that of the incompressible plane channel.
However, a quantitative comparison with the channel flow is close to impossible, 
owing to additional curvature effects, the presence of pressure gradients, 
and the unavoidable presence of spatial transients.

Through the analysis of the temporal transient after the imposition of wall forcing, we have described how the localized reduction of friction on the suction side of the wing alters the interaction between the shock wave and the turbulent boundary layer.
While the response of wall friction to the wall-based actuation is quite fast, 
the evolution of the flow from the uncontrolled state to the quasi-stationary 
conditions with control is significantly longer.
The initial local reduction in friction perturbs the state of equilibrium of the flow, 
eventually resulting in the downstream displacement of the shock.
The process is relatively slow, taking place on a time scale of about eight 
convective time units.
When the control parameters are optimally tuned, the downstream shift of the shock 
is significant, and the shock becomes stronger: the turbulent boundary layer undergoes 
a stronger adverse pressure gradient and separates.
In these cases, a small recirculation bubble, which may alter the topology of the 
shock system, appears beneath the shock.
The length of the detached region, 
when present, is found to directly correlate with the values of friction drag reduction.
Further works should explore the effects of changing Mach numbers on the aerodynamic performance of the controlled wing, to establish whether the improvements are only due to the shift (and the presence) of the shock wave.

The main limitation of this work consists in the low value of the Reynolds number, 
namely $\Rey_\infty = 3 \times 10^5$, which is not high enough to be representative of the typical flight conditions, 
and leads to a subcritical turbulent boundary layer on a fraction of the wing. 
The numerical tripping required to guarantee a reasonable development 
of turbulence introduces an extra layer of arbitrariness. 
We believe that considering a flight Reynolds numbers of at least $Re_\infty=10^6$, 
is required to provide the observations described in this work with a firmer 
physical ground.
Nevertheless, drag reduction in the order of that observed in this study would represent 
a large improvement in terms of operational cost of transonic flight, and definitely motivates further research in the field of turbulent drag reduction in realistic flow configurations.

As a closing remark, we underline that the present conclusions should not be limited to spanwise forcing, as they naturally extend to any drag-reducing technology, including passive devices such as riblets, which are much closer than StTW to industrial implementation.
More than optimizing the StTW control parameters for this specific wing, future work should properly formulate and solve an optimization problem for the optimal layout and positioning for an actuator, or, equivalently, for a passive device for skin friction reduction.
Since at least part of the cost for (active and passive) flow control directly relates to the covered surface area, whereas the benefit, as we have seen here, crucially depends on which area is selected, the designer will need to know how to deploy the control system optimally to achieve the best cost/benefit ratio.

\appendix
\section{How to define viscous scaling}
\label{app:friction}

\begin{figure}
\centering
\includegraphics{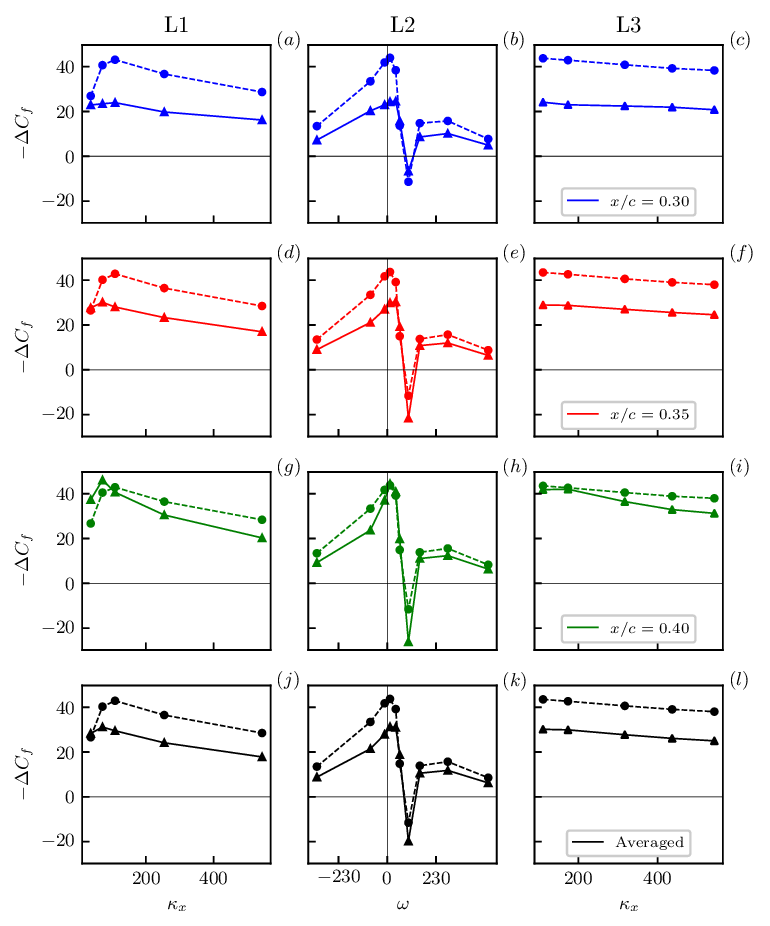}
\caption{Dependence of $\DCf$ upon the measurement position along the chord. Circles and dashed lines: channel flow data; triangles and solid lines: present data.}
\label{fig:appendixChannelvsTragfoil}
\end{figure}

As mentioned in \S\ref{sec:parameters}, a true comparison between the skin-friction drag-reducing properties of StTW in the incompressible plane channel flow and the present transonic wing is unavoidably arbitrary to some extent.
In this Appendix, we critically discuss our choice of extracting the friction velocity required to define viscous scaling at the specific position $x=0.4$, and show the effect of alternative choices. 
In particular, we consider the positions $x = 0.3, x = 0.35, x = 0.4$, as well as the option of obtaining $\DCf$ by averaging over the range $0.3 \le x \le 0.4$.
All these positions are located at least $10\%$ downstream of the beginning of the control $x_b$, and at the same time are sufficiently upstream of the shock position that the influence of the latter can be reasonably neglected.

Figure \ref{fig:appendixChannelvsTragfoil} shows the values of $\DCf$ obtained from the present dataset, together with results interpolated from the incompressible channel flow data by \cite{gatti-quadrio-2016}, for the same control input.
From a global perspective, the results closest to the channel flow data are obtained in the central and right-most panels, corresponding to lines L2 and L3 of figure \ref{fig:paramspace}, respectively.
In panels $(a), (d), (g)$ and $(j)$, corresponding to line 1 of figure \ref{fig:paramspace}, the position of the peak of $\DCf$ is shifted at lower frequencies for the wing, albeit the two curves are qualitatively similar.
The obvious deduction is that the more downstream $C_f$ is extracted, the more the quantity $\DCf$ tends to the value of the incompressible channel flow, regardless of the control parameters. 
In particular, the location $x=0.3$, whose results are depicted in panels $(a), (b)$ and $(c)$ is probably still too close to the beginning of the actuation, while the results closest to the channel flow data are those for $x=0.4$ (panels $(g), (h)$ and $(i)$), i.e., where $\Delta C_f(x)$ is reaching a plateau.


\begin{figure}
\centering
\includegraphics{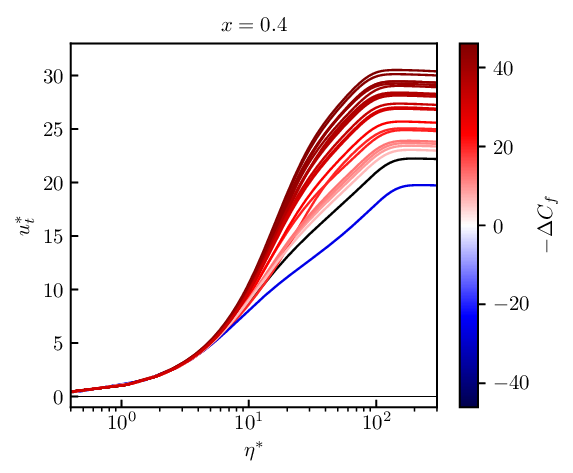}
\caption{Wall-normal profile of the wall-parallel velocity component $u_t$, scaled in actual viscous units. The profiles for simulations C1--C20 (constant forcing amplitude $A$) are extracted at $x=0.4$, and coloured according to the skin-friction reduction rate $-\Delta C_f$. The black line denotes the reference case REF.}
\label{fig:VelProf}
\end{figure}

By using actual viscous units computed with friction velocity at $x=0.4$, figure \ref{fig:VelProf} shows the wall-normal profiles of the mean streamwise velocity component, plotted according to the Trettel--Larsson scaling \citep{trettel-larsson-2016}, and coloured according to the skin-friction reduction rate $\Delta C_f$.
The profiles provide information on what would be the true Reynolds-independent measure of the skin-friction reduction in an indefinite plane channel, i.e. the wall-normal shift of the logarithmic layer of the streamwise velocity component \cite{gatti-quadrio-2016}.
It should be remarked, however, that the concept of logarithmic law should be used with caution here, as the flow has a rather low $Re$ and is out of equilibrium.
With this scaling, the viscous sublayer is obviously not affected by StTW, whose effect is instead visible in the logarithmic region, where the curves are shifted towards higher values of $u^\ast$ for a fixed $\eta^\ast$ proportionally to the skin-friction reduction rate.
The only curve below the reference one corresponds to case DI, in which $C_f$ is indeed increased.
The low Reynolds number of the simulations of this study, an issue mentioned in \S\ref{sec:conclusions}, is reflected in figure \ref{fig:VelProf} in the small range of wall-normal distances $\eta^\ast$ in which the scaled streamwise velocity profile follows a linear law in logarithmic axis.
This range, indeed, is similar to that observed in the study by \cite{gatti-quadrio-2016} in their figures 10a and 10b, which show data at the lowest friction Reynolds number available ($\Rey_\tau=200$).

\begin{figure}
\centering
\includegraphics[width=\textwidth]{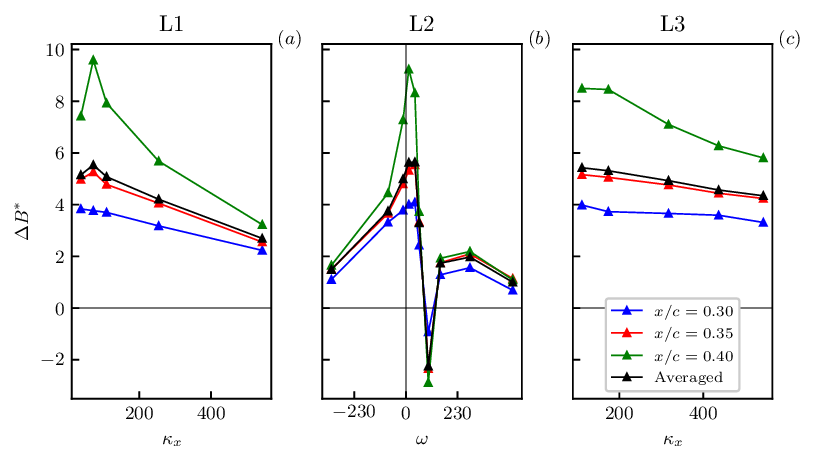}
\caption{Wall-normal shift of the logarithmic layer of the streamwise velocity component for the simulations at constant $A$. The different curves denote different streamwise locations for extracting the data.}
\label{fig:AppendixDeltaB}
\end{figure}

Finally, figure \ref{fig:AppendixDeltaB} shows the wall-normal shift $\Delta B^\ast$ extracted at $\eta^\ast=90$ for the set of simulations at constant forcing amplitude, computed on different stations of the suction side of the wing.
Additionally, the black curves show $\Delta B^\ast$ obtained considering the velocity profile averaged in the range $0.3\le x\le 0.4$.
The qualitative behaviour confirms the observations made in reference to figure \ref{fig:appendixChannelvsTragfoil} for the skin-friction coefficient, confirming that the two quantities are related, and that the conclusions by \cite{gatti-quadrio-2016} also apply to compressible boundary layers over mildly non-planar surfaces.



\section*{Acknowledgments}
All the simulations were performed on the Hawk Cluster in Stuttgart (Baden-Württemberg, Germany), within the project TuCoWi. The authors are grateful to Dr Alessandro Chiarini and Dr Antonio Memmolo for their help in the initial stages of the work.

\section*{Funding} 
This research received no specific grant from any funding agency, commercial or not-for-profit sectors.

\section*{Declaration of Interests} 
The authors report no conflict of interest.

\section*{Author ORCIDs}
Niccolò Berizzi, https://orcid.org/0009-0006-5570-0014\\
Davide Gatti, https://orcid.org/0000-0002-8178-9626\\
Giulio Soldati, https://orcid.org/0009-0007-4597-756X\\
Sergio Pirozzoli, https://orcid.org/0000-0002-7160-3023\\
Maurizio Quadrio, https://orcid.org/0000-0002-7662-3576

\bibliographystyle{jfm}
\bibliography{../Wallturb.bib}

\end{document}